\documentclass[10pt,journal,compsoc]{IEEEtran}

\ifCLASSOPTIONcompsoc
  \usepackage[nocompress]{cite}
\else
  \usepackage{cite}
\fi

\def\FIGDIR{./figures}

\usepackage{amsmath,amssymb,amsfonts}
\usepackage{color}
\usepackage{textcomp}

\usepackage[switch]{lineno}

\usepackage[ruled,linesnumbered,noend]{algorithm2e}
\SetKwComment{Comment}{$\triangleright$\ }{}
\SetAlFnt{\small}

\SetCommentSty{mycommfont}
\usepackage[usenames,dvipsnames,table,xcdraw]{xcolor}
\usepackage{mathtools}
\usepackage{diagbox}
\usepackage{subcaption}

\makeatletter
\def\thickhline{%
  \noalign{\ifnum0=`}\fi\hrule \@height \thickarrayrulewidth \futurelet
   \reserved@a\@xthickhline}
\def\@xthickhline{\ifx\reserved@a\thickhline
               \vskip\doublerulesep
               \vskip-\thickarrayrulewidth
             \fi
      \ifnum0=`{\fi}}
\makeatother
\newlength{\thickarrayrulewidth}
\usepackage{setspace}
\usepackage{comment}
\usepackage[most]{tcolorbox}
\usepackage{pifont}

\usepackage{endnotes,xspace,graphicx,fancyvrb,multirow}
\usepackage{multicol}
\usepackage{paralist}

\usepackage{mathptmx} % This is Times font

\newcommand{\ignore}[1]{}
\usepackage{fancyhdr}
\usepackage[normalem]{ulem}
\usepackage[hyphens]{url}
\usepackage{microtype}
\usepackage{flushend}
\usepackage{hyperref}

\usepackage{enumitem}
\usepackage{booktabs} % for professional tables

\newcommand{\PCignore}[1]{}

\newcommand{\insertFigure}[2]{
    \begin{figure}[t]
\setlength{\abovecaptionskip}{-3pt}
\setlength{\belowcaptionskip}{-1pt}
        \centering
        \includegraphics[width=\linewidth]{\FIGDIR/#1.pdf}
	\vspace{-2mm}
        \caption{#2}
        \label{fig:#1}
    \end{figure}
}

\newcommand{\insertWideFigure}[2]{

    \begin{figure*}[ht]
    \setlength{\abovecaptionskip}{-2pt}
    \setlength{\belowcaptionskip}{-4pt}
        \centering
        \includegraphics[width=\textwidth]{\FIGDIR/#1.pdf}
	    \vspace{-2mm}
        \caption{#2}
        \label{fig:#1}
    \end{figure*}
}

\newcommand{\squishlist}{
 \begin{list}{$\bullet$}
  { \setlength{\itemsep}{0pt}
     \setlength{\parsep}{3pt}
     \setlength{\topsep}{3pt}
     \setlength{\partopsep}{0pt}
     \setlength{\leftmargin}{1.5em}
     \setlength{\labelwidth}{1em}
     \setlength{\labelsep}{0.5em} } }

\newcommand{\squishlisttwo}{
 \begin{list}{$\bullet$}
  { \setlength{\itemsep}{0pt}
     \setlength{\parsep}{0pt}
    \setlength{\topsep}{0pt}
    \setlength{\partopsep}{0pt}
    \setlength{\leftmargin}{2em}
    \setlength{\labelwidth}{1.5em}
    \setlength{\labelsep}{0.5em} } }

\newcommand{\squishend}{
  \end{list}  }

\newcommand{\betterparagraph}[1]{\noindent\textbf{#1.}}

\newcommand{\colonparagraph}[1]{\noindent\textbf{#1:}}

\newcommand{\newmaestro}{MAESTRO-BLAS\xspace}
\newcommand{\maestro}{MAESTRO\xspace}
\newcommand{\gemm}{GEMM\xspace}
\newcommand{\blas}{BLAS\xspace}
\newcommand{\flash}{FLASH\xspace}

\begin{document}
\title{Evaluating Spatial Accelerator Architectures with Tiled Matrix-Matrix Multiplication}

\author{Gordon~E.~Moon\IEEEauthorrefmark{1},
        Hyoukjun~Kwon\IEEEauthorrefmark{2},
        Geonhwa~Jeong\IEEEauthorrefmark{2},
        Prasanth~Chatarasi\IEEEauthorrefmark{2},
        Sivasankaran~Rajamanickam \IEEEauthorrefmark{3},
        and~Tushar~Krishna\IEEEauthorrefmark{4}% <-this % stops a space
\IEEEcompsocitemizethanks{
\IEEEcompsocthanksitem
\IEEEauthorrefmark{1} Department of Software, Korea Aerospace University, Gyeonggi, Goyang, Republic of Korea (ehmoon@kau.ac.kr),
% \protect\\
% E-mail: see http://www.michaelshell.org/contact.html
\IEEEcompsocthanksitem 
\IEEEauthorrefmark{2} School of Computer Science, Georgia Institute of Technology, Atlanta, GA, 30332 (hyoukjun, geonhwa.jeong, cprasanth @gatech.edu).
%\IEEEcompsocthanksitem G. Jeong is with the School
%of Computer Science, Georgia Institute of Technology, Atlanta,
%GA, 30332 (email: geonhwa.jeong@gatech.edu).
\IEEEcompsocthanksitem 
\IEEEauthorrefmark{3} Center for Computing Research, Sandia National Laboratories, Albuquerque, NM, 87123 (srajama@sandia.gov). Sandia National Laboratories is a multimission laboratory managed and operated by National Technology \& Engineering Solutions of Sandia, LLC, a wholly owned subsidiary of Honeywell International Inc., for the U.S. Department of Energy’s National Nuclear Security Administration under contract DE-NA0003525.
\IEEEcompsocthanksitem
\IEEEauthorrefmark{4} School of Electrical and Computer Engineering, Georgia Institute of Technology, Atlanta,
GA, 30332 (tushar@ece.gatech.edu).}% <-this % stops an unwanted space
\thanks{}}

% Manuscript received February 24, 2021; revised May 5, 2021.


% Submission to IEEE Transactions on Parallel and Distributed Systems

\markboth{}%
{Moon \MakeLowercase{\textit{et al.}}: Evaluating Spatial Accelerator Architectures with Tiled Matrix-Matrix Multiplication}



\IEEEtitleabstractindextext{%
\begin{abstract}
There is a growing interest in custom spatial accelerators for machine learning applications. 
These accelerators employ a spatial array of processing elements (PEs) interacting via custom buffer hierarchies and networks-on-chip. 
%The focus of this work is to evaluate such novel accelerators using a tiled general matrix-matrix multiplication (\gemm) kernel.
The efficiency of these accelerators comes from employing optimized dataflow (i.e., spatial/temporal partitioning of data across the PEs and fine-grained scheduling) strategies to optimize data reuse.
%This enables them to support an extremely large set of mappings compared to accelerators like GPUs.
%
%\gemm is a key kernel in both computational science applications and machine learning applications.
The focus of this work is to 
evaluate these accelerator architectures using a tiled general matrix-matrix multiplication (\gemm) kernel.
To do so, we develop
%In order to evaluate these architectures thoroughly, 
%we propose 
a framework that finds optimized mappings (dataflow and tile sizes) for a tiled \gemm for a
given spatial accelerator and workload combination, leveraging 
an analytical cost model for runtime and energy.
%and estimate runtime and energy
%using a detailed
%analytical model.
%by using an analytical model that could accurately model the spatial accelerators.
%Our framework comprises of a mapping explorer that can elaborate all possible mapping choices (dataflow and tiles sizes) for spatial accelerators, and prune the space based on hardware constraints and a heuristic for tile sizes that aims to
%maximize scratchpad occupancy.
Our evaluations over five spatial accelerators 
%using the analytical model 
demonstrate that the tiled \gemm %dataflows 
mappings
systematically 
generated by our framework achieve high performance on various \gemm 
workloads and accelerators. %Experimental results compare the runtime and energy costs for each tiled \gemm algorithm, workload, and accelerator.
% Our evaluations over five spatial accelerators demonstrate \gemm mappings constrained by accelerators and systematically generated by our mapping explorer achieve the optimal performance on various \gemm workloads and accelerator configurations.
% In this work, we study a suite of mappings (the dataflow and tile sizes) using a detailed analytical model \newmaestro.
% %inspired by current custom accelerators and potential future accelerators. 
% \newmaestro allows evaluating mappings of \blas operations on current and potential future accelerators. %accurately. 
% We study the general matrix-matrix multiplication kernel (\gemm) which could be expressed with different mappings. We present a comprehensive study of different mappings for \gemm. In addition, we develop a tiling scheme that is suitable for a custom accelerator. The study shows the advantages of flexible accelerators when required to support different mappings for \gemm. %\textcolor{red}{(150 / 150 words)}
\end{abstract}

% old
%As the line between machine learning and computational science gets blurred these custom accelerators have the potential to be part of a heterogeneous node architecture on Top500 systems.
%The use of these accelerators for computational science remains an open question.
% 2nd
%The use of spatial accelerators for computational science remains unexplored and is the focus of this work.
%an open question, that this work seeks to address.
%
% Since linear algebra kernels for computational science heavily rely on tiled general matrix-matrix multiplication (\gemm),
%algorithms for data reuse.

}

\maketitle

\IEEEdisplaynontitleabstractindextext

% \fi
\IEEEpeerreviewmaketitle

\IEEEraisesectionheading{
\section{Introduction}\label{sec:introduction}}
% \GM{Although several CONV-specialized spatial accelerators have been developed so far, searching the optimal mapping space of CONV still remains an open question. And everyone knows the GEMM can be considered as a special case of CONV. Explain why exploring the mapping space of even more simplified GEMM than CONV is an important problem on spatial accelerators.}
% \GM{Many reviewers were curious about why we explore GEMM rather than CONV. They are thinking GEMM is a subset of CONV. They might think we addressed a simpler GEMM mapping problem compared to exploring CONV mapping. Need to precisely define our problem (flattening CONV (e.g, TPU? GEMM) vs. direct CONV (cite CPU/GPU paper) Most of the CPU/GPU/spatial accelerators directly perform CONV without flattening)}
% \PC{All my comments are based on the observation that PPoPP audience are experienced with only CPUs/GPUs, and very less on any other architectures, especially ML accelerators.}

\IEEEPARstart{S}{everal} custom ASIC accelerators have emerged in the recent past and 
have been successfully used to exploit massive parallelism and locality in the machine learning (ML) applications.
The most popular examples are systolic arrays such as TPU~\cite{jouppi2017datacenter}, xDNN~\cite{xDNN-web}, RAPID~\cite{DBLP:conf/vlsic/FleischerSZSOSC18} and advanced forms such as NVDLA~\cite{nvdla},  Eyeriss~\cite{chen2016eyeriss},  ShiDianNao~\cite{du2015shidiannao} and MAERI~\cite{kwon2018maeri}.
Accelerators such as these are already being integrated in Petascale systems. For example, the Lassen system at Lawrence Livermore National Labs has a Cerebras accelerator integrated with it. 
Recent results show a 0.86 PFLOPS on a single wafer scale chip \cite{rocki2020fast} on stencil problems. Graphcore IPUs and SambaNova are starting to be use in traditional HPC applications \cite{louw2021using,emani2021accelerating}. The advantages provided by these accelerators vary from reduced data movement due to the data flow on chip and opportunities to accelerate new applications that were not amenable to accelerators like GPUs. These benefits and recent successes demonstrate potential for these accelerators to be part of future exascale systems.

As more such heterogeneous systems are expected in the exascale era, it is important to develop a methodology for modeling such accelerators.
These accelerators have demonstrated lower runtime and higher energy efficiency relative to existing popular architectures such as multi-core CPUs and many-core GPUs~\cite{jouppi2017datacenter}.
%The success of these accelerators within the context of ML raises the question if some variants of these accelerators could also be used for computational science and engineering (CSE) applications, making it an ideal candidate for integration into a heterogeneous HPC node alongside CPUs and GPUs.
%with %comprising 
%CPUs, GPUs, and accelerators.
%Such a custom accelerator could be useful for both machine learning applications and computational science applications, making it an ideal candidate for integration into a heterogeneous HPC node with CPUs, GPUs, and accelerators.
%system %. with custom accelerators and traditional hardware (CPU/GPU).
%\TK{@Siva - do the following lines seem distinguishing accelerators from CPUs and GPUs seem fine? I made some edits to the original text}
%\SR{@Tushar - yes, this looks good.}
The primary architectural features that distinguish these ``spatial'' accelerators for ML from CPUs and GPUs are parallelism using hundreds to thousands of processing elements (\textbf{PEs}), a fast network-on-chip \textbf{(NoC)} connecting these and use of private/shared scratchpad buffers for data reuse. Different accelerators differ in their  \textbf{dataflow}, i.e., mechanism for data reuse over buffers and wires.
\gemm (General Matrix-Matrix Multiplication) is a key computational kernel within ML, computational science and engineering (CSE) applications, and other computational kernels~\cite{kaagstrom1998gemm,goto2008high}. 
%level 3 \blas kernels such as LU factorization, triangular matrix-matrix multiply, triangular-solve with multiple right-hand sides, symmetric rank-$k$ update, and Cholesky QR factorization~\cite{kaagstrom1998gemm,goto2008high,anderson1999lapack, yamazaki2015mixed}. 
%Recent work analyzed tiling and batching for \gemm on GPUs \cite{li2019coordinated}. 
In this work, we focus
on modeling and evaluating multiple spatial accelerator architectures using this important kernel.
%\gemm is an ideal kernel for such a study in some sense as it could be thought of as a special case of the ML use cases these accelerators are designed to address and important for \cse applications. 
%However, 
%as we will show, the performance of spatial accelerators can vary drastically even 
% for a simple kernel like \gemm 
% depending on
% the workload dimensions, 
% accelerator's dataflow,
% and buffer/NoC 
% constraints.
%
%\SR{Need Better example} For example, a dataflow choice of parallelizing across channels employed by NVDLA to target DNNs like ResNet50 may not be efficient for \gemm.
%
%directly applicable for a CSE use case.
%
%behavior could be different based on workloads even for a kernel like \gemm. The hardware constraints developed for ML use cases are not the best choices for CSE use cases. For example, parallelizing channels may not be directly applicable for a CSE use case. 
%
% In this work, we develop an optimizer to 
% find optimized mappings (defined based on projected runtime) of \gemm over \emph{multiple} spatial accelerators.
\gemm kernel has been studied heavily on traditional architectures in the context of CSE and ML workloads.  Evaluating all the algorithm choices for \gemm, while considering workload dimensions, tiling strategies etc. on even a single new spatial accelerator is a challenge. We aim to do this evaluation on five different accelerators (via simulation).
%\TK{I added via simulation so reviewers dont think we ran on actual systems}
% SR :I agree.
Optimizing the \gemm mapping on 
spatial accelerators poses the following challenges.
(i) the sizes of input matrices vary significantly;
(ii) the space of possible mappings involving multi-level tiling, parallelization, and loop orders could be in the billions (see Section~\ref{subsec:search_space_pruning} for an example);
% \footnote{\textcolor{red}{Will add a footnote on how we calculated this, or fwd point to some section.}}
(iii) the dataflow choice of the accelerator and its internal structures (e.g., multi-cast, reduction tree) have to be accounted for as constraints (e.g., \autoref{table:flash_mapping_comparison} summarizes these restrictions %and what is possible 
in five ML accelerators);
%(3)  wide variety of hardware structures (e.g., multi-cast, reduction tree) in the accelerators have to be accounted for, and
(iv) accurate cost models to model diverse accelerator dataflows and estimate performance %statistics 
of a mapping on the accelerator are needed.
\insertWideFigure{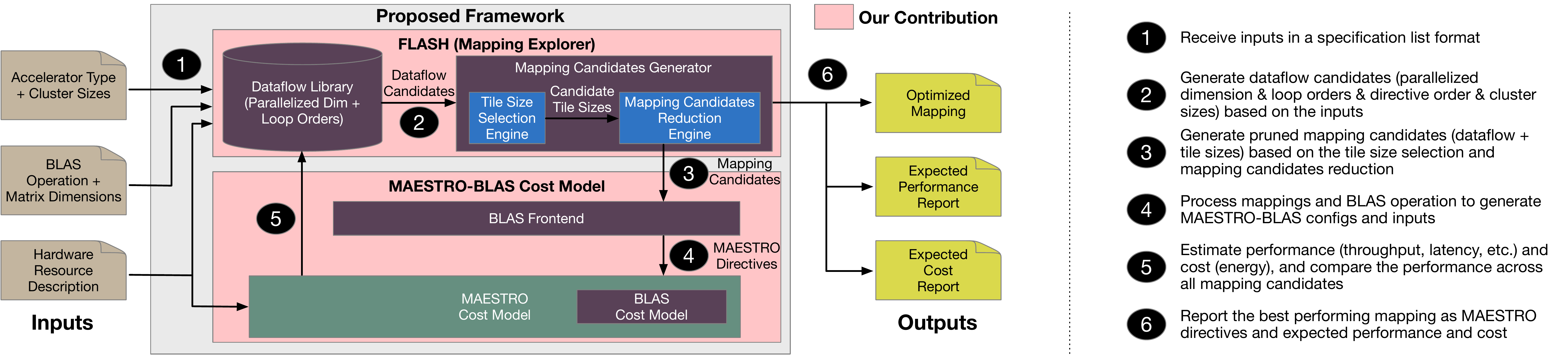}{An overview of the proposed framework for searching for optimal mappings with walk-through of steps.}

To address the challenges highlighted above and evaluate the different accelerators, we take a systematic approach. First, we develop a \emph{methodology for modeling accelerators} for the tiled \gemm kernel (\autoref{sec:target_accelerators}) which could be extended to other kernels in the future. Next, we develop a framework shown in \autoref{fig:FLASH_Overview_wide} which is the \emph{second contribution} of this work. Our framework
finds the best mappings and tile sizes (defined based on projected runtime) of tiled \gemm over \emph{multiple} spatial accelerators using an accurate analytical model.
Our framework is comprised of two key components.
(1) a mapping explorer \textbf{\flash} (Flexible Linear Algebra dataflow via Spatio-temporal Hierarchical-mapping) that can explore a high dimensional space of tile sizes, loop ordering choices, parallelization, by pruning the search space given a workload, hardware description and  dataflow constraints (addressing challenges (i)-(iii) above);
%\flash prunes the search space of mapping choices using dataflow constraints of accelerators, analytically derived range of tile sizes based on hardware description and problem dimensions. 
(2) an analytical cost model \textbf{\newmaestro} that can be used to 
evaluate the accelerators for mappings generated by \flash (addressing challenge (iv)). 

The \emph{third contribution} of this work is a thorough experimental evaluation of the dataflows and tile sizes for six representative \gemm workloads on five popular spatial accelerators (which offer varying levels of flexibility in their dataflow) and two accelerator configurations (cloud and edge). We also evaluate the performance of the five accelerators on Deep Neural Networks (DNNs) where \gemm kernel is foundational to training and inference.
%We use \flash to optimize \gemm mappings on five popular spatial accelerators which offer varying levels of flexibility in their dataflow.
%that are either rigid or {\em flexible} in their data orchestration for a given workload. 
%We make three primary contributions in this work.
%First, we present \textbf{\newmaestro}, a cost model for evaluating the performance and energy implications of different mappings of \blas kernels on spatial accelerators. \newmaestro is built upon \maestro~\cite{kwon2019understanding}, an open-source analytical modeling tool for evaluating the performance of running DNNs on spatial accelerators. 
%Second, we propose a suite of optimized mappings for \gemm called \textbf{\flash} (Flexible Linear Algebra dataflow via Spatio-temporal Hierarchical-mapping). \flash includes various dataflow strategies (including some inspired by current accelerators), and an analytically determined optimal multi-level tiling strategy for maximizing reuse.
%Second, we use an analytically determined optimal multi-level tiling strategy for maximizing reuse and the hardware constraints to reduces the search space for best mapping in \textbf{\flash} (Flexible Linear Algebra dataflow via Spatio-temporal Hierarchical-mapping).

These contributions result in significant new results.
The novel pruning approach in our framework reduces the search space by 99.7\% for a  (256$\times$256)$\times$(256$\times$256) \gemm and still finds a correct mapping. 
%We use \newmaestro on the reduced search space to arrive at the best mapping of \gemm for an accelerator.
This also leads to decreasing the time for searching through the space using \newmaestro by 99.9\%.
We derive the tile sizes for \gemm on spatial accelerators analytically and show the tile sizes chosen by \flash reduces runtime up to 94\% and energy by up to 96\%.
The experimental evaluation using the analytical model shows the importance of loop order, matrix shape,
and how flexible dataflow accelerators (e.g., MAERI~\cite{kwon2018maeri}), can provide significant runtime and energy improvements as opposed to fixed dataflow accelerators.
The novel coupling of search space pruning that accounts for hardware, workload, and algorithm choice with an analytical model is a generalizable framework for other workloads and architectures in the future. 

\section{Background}
\label{sec:background}

% \TK{There seems to be a lot of repetition in Sec II-B. I think its because text that was earlier in Intro has now moved here?}
% \GM{Yes. It is. We have to significantly shrink overall Section 2-B. I think Section 2-B can be summarized in one column of page including Figure 2. Then we can reduce one page of the paper.}

\subsection{General Matrix-Matrix Multiplication (\gemm)}
\label{subsec:background_gemm}
\gemm kernel multiplies two input matrices \textit{A} of size \textit{M}$\times$\textit{K} and \textit{B} of size \textit{K}$\times$\textit{N} to obtain an output matrix \textit{C} of size \textit{M}$\times$\textit{N}, as shown in Algorithm \ref{algorithm:GEMM}.
% as shown in \autoref{fig:background_mapping}.
% as shown in Algorithm \ref{algorithm:GEMM}. 
%
\gemm performs \textit{M}$\times$\textit{N}$\times$\textit{K} MACs (Multiply-ACcumulate operations) within three nested loops. The order of three loops can be changed to exploit various data reuses of matrices \textit{A}, \textit{B} and \textit{C} across space and time.
\begin{algorithm}
\SetAlgoLined
\KwIn{\textit{A}[\textit{M}][\textit{K}], \textit{B}[\textit{K}][\textit{N}]}
\KwOut{\textit{C}[\textit{M}][\textit{N}]}
 \For{\textit{m} $=$ $0$ \KwTo \textit{M}$-$$1$}{
    \For{\textit{n} $=$ $0$ \KwTo \textit{N}$-$$1$}{
        \For{\textit{k} $=$ $0$ \KwTo \textit{K}$-$$1$}{
            \textit{C}[\textit{m}][\textit{n}] $+$$=$ \textit{A}[\textit{m}][\textit{k}] $\times$ \textit{B}[\textit{k}][\textit{n}]
        }
    }
 }
 \caption{General Matrix-Matrix Multiplication}
 \label{algorithm:GEMM}
\end{algorithm}

% \begin{algorithm}
% \SetAlgoLined
% \KwIn{matrix \textit{A} of size (\textit{M}$\times$\textit{K}) and matrix \textit{B} of size (\textit{K}$\times$\textit{N})}
% \KwOut{matrix \textit{C} of size (\textit{M}$\times$\textit{N})}
%  \For{\textit{m} $=$ 0 \KwTo \textit{M}$-$1}{
%     \For{\textit{n} $=$ 0 \KwTo \textit{N}$-$1}{
%         \For{\textit{k} $=$ 0 \KwTo \textit{K}$-$1}{
%             \textit{C}[\textit{m}][\textit{n}] += \textit{A}[\textit{m}][\textit{k}] $\times$ \textit{B}[\textit{k}][\textit{n}]
%         }
%     }
%  }
%  \caption{General Matrix-Matrix Multiplication}
%  \label{algorithm:GEMM}
% \end{algorithm}

Several state-of-the-art Deep Neural Network models %such as multi-layer perceptron, recurrent neural networks, and convolutional neural networks 
spend a large fraction of the training/inference time on \gemm operations~\cite{sigma_hpca2020}. 
%\gemm operations are also foundational to several CSE applications such as boundary element methods in electromagnetic simulations~\cite{langston2018massively}.
%\TK{The following line is not very clear. Is GEMM the key kernel within LU and QR factorization, or is GEMM a key kernel along with other level 3 \blas kernels. Since we only focus on GEMM, I dont think we need the line about other BLAS-3 kernels}. 
In addition, \gemm is the key computational kernel in CSE applications,  level 3 \blas kernels such as LU factorization, triangular matrix-matrix multiply, triangular-solve~\cite{kaagstrom1998gemm,goto2008high},
%\gemm is also an important part of
sparse direct methods for LU/QR factorizations~\cite{davis2016survey}, 
%\gemm is also useful for 
and sparse iterative solvers~\cite{bavier1970amesos2}. 
There are also smaller batched \gemm kernels that are critical for multiphysics codes~\cite{abdelfattah2016performance,howard2018employing,kim2017designing}.
%\TK{Can we lots of citatations for CSE kernels that use GEMMs?}
Thus, addressing the performance of \gemm kernel would have a broad impact across CSE and ML applications. 
% Moreover, the performance of \gemm is also a key step in evaluating the suitability a new accelerators for CSE and ML applications.
%This has also been the necessary (not sufficient) step of evaluating the suitability a new architecture for CSE applications.
The primary difference between all these use cases is the size and shape of input matrices for \gemm. Experiments in this paper vary the size and shape of matrices, tile sizes and loop order to cover all these use cases. 
%We consider non-tiled and tiled \gemm kernels with different loop orders that are well studied in other architectures. 
\emph{The key difference that emerges in this study is that while one loop order is typically the best among GPUs (or CPUs) even from different vendors, the same is not true for spatial accelerators as they are quite diverse.}
\insertFigure{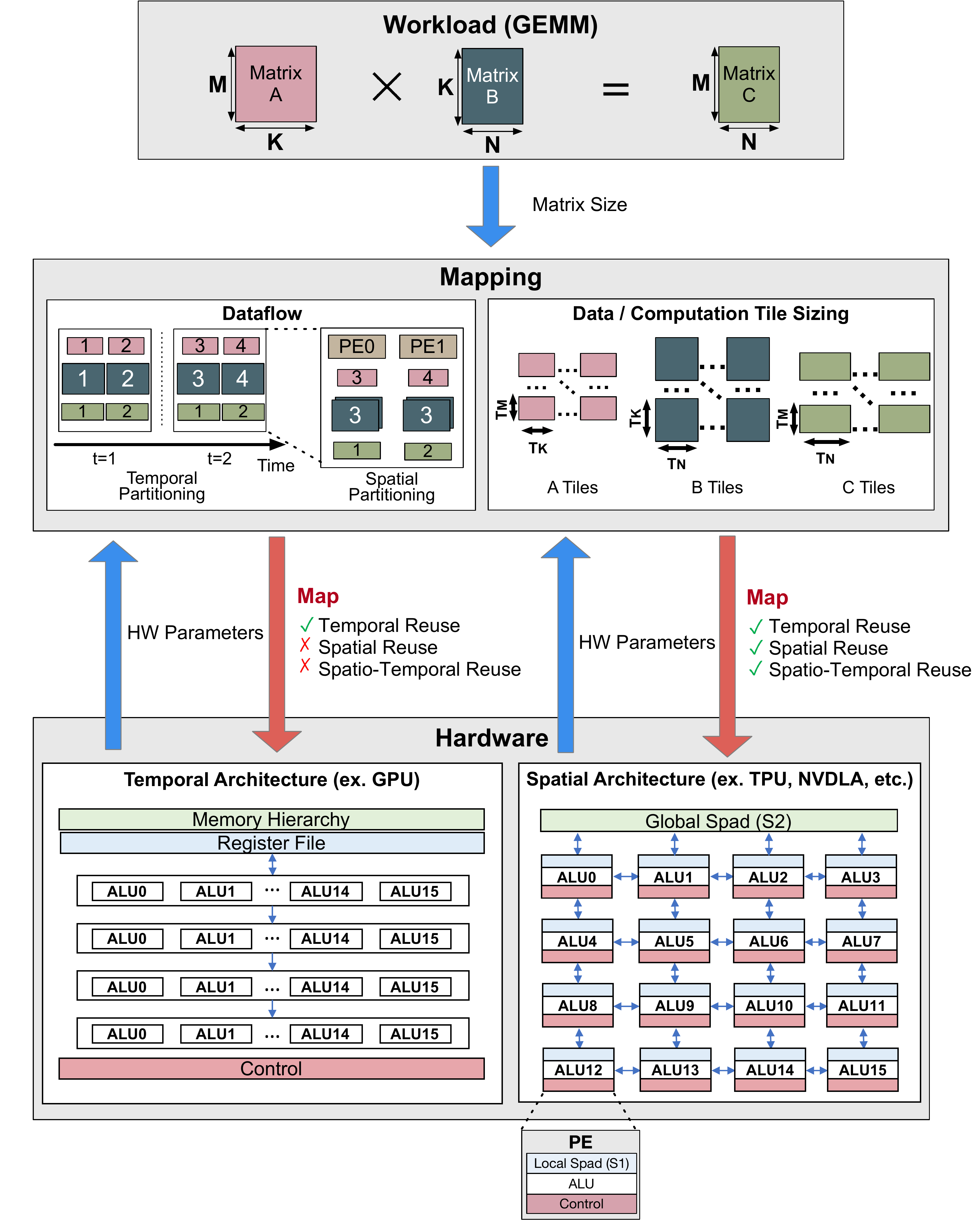}{An overview of mapping a \gemm workload on temporal architecture such as GPU, and spatial architecture such as TPU and NVLDA. \gemm mapping (middle) shows how the data for \gemm (top) could be partitioned (middle left) temporally or spatially and appropriate tile size are chose (middle right).
%Spatial architecture has an array of Processing Elements as shown in right-bottom corner, which are specialized to support specific workloads such as DNNs. Mapping of a workload on an architecture refer to tile sizing of the workload and temporal/spatial partitioning of tiles.
% The latter is termed as Dataflow~\cite{chen2016eyeriss}. 
%$\textit{T}_{\textit{M}}$, $\textit{T}_{\textit{N}}$ and $\textit{T}_{\textit{K}}$ refer to the tile sizes for \textit{M}, \textit{N} and \textit{K} dimensions, respectively.
}

\subsection{Spatial Accelerators}
\label{subsec:background_spatial_accelerators}

\betterparagraph{Architecture}
\autoref{fig:background_mapping} shows an abstract spatial accelerator architecture, which consists of a processing element (PE) array, a global shared scratchpad memory (called S2), and a controller.
% \SR{can we highlight spatial accelerator in the figure with a bold box/color?}
This template is common across all state-of-the-art ASIC accelerators~\cite{chen2016eyeriss, jouppi2017datacenter,kwon2018maeri,nvdla, du2015shidiannao}. Table \ref{table:target_accelerators} lists the ones we consider in this work.
Network-on-chips (NoCs)
%\footnote{We refer any form of on-chip interconnects among PEs as NoC (neighbor-to-neighbor, bus, etc.), not limited to mesh topology-based NoCs.} 
interconnect the PEs and S2 global scratchpad memory.
PEs are the compute units of accelerators that contain ALUs or Multiply-Accumulate 
units
with a small local scratchpad (called S1). Like a cache memory, scratchpad memory reduces the number of remote buffer accesses.
However, unlike a cache memory, data layout and insertion/eviction of data to a scratchpad memory are fully customizable for programmers or accelerator designers.
%, which enables application specific optimizations in the context of specialization.
 Spatial accelerators exploit not only global buffer-PE communication but also inter-PE communication patterns~\cite{kwon2017rethinking} (i.e., dataflow among PEs and global buffer).
The inter-PE communication pattern could be flexible or fixed. Fixed inter-PE communication pattern restricts which (\gemm) algorithm could be run on an accelerator. \emph{Our framework takes into account the fixed/flexible communication patterns in addition to the typical architecture features such as number of PEs and S1/S2 size to faithfully simulate different accelerators}.

%The communication patterns are determined by the \textit{mapping} of the workload on the accelerator, which we define later in this section.
%
%, which refers to the dataflow strategy and the data tile sizes, as we define later in this section.

\insertFigure{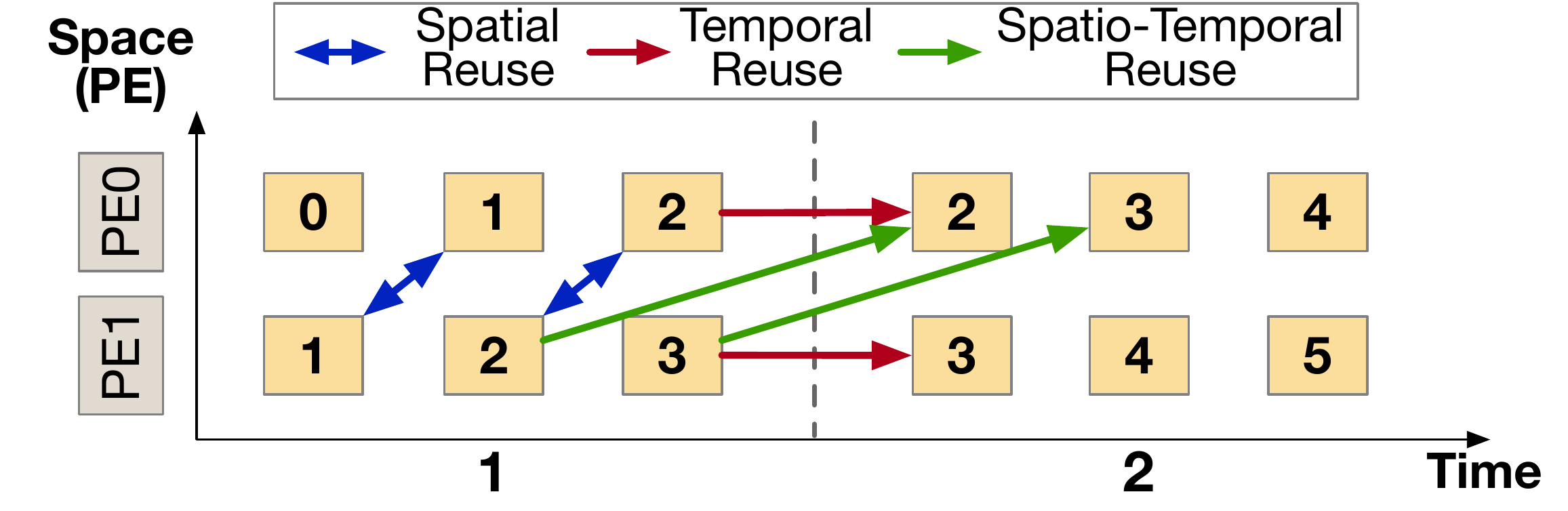}{Data reuse in the example mapping on two PEs. Spatial and spatial-temporal reuse are leveraged via the NoC wires while temporal reuse is leveraged via scratchpad buffers. Yellow boxes and numbers inside them represent data and data indices, respectively.}

\betterparagraph{Data Reuse in Spatial Accelerators}
Efficiency in spatial accelerators comes from  exploiting data reuse.
Data reuse in spatial accelerators can be categorized into three types using the temporal and spatial nature~\cite{kwon2019understanding}.
~\autoref{fig:background_datareuse} provides an example mapping and data reuse analysis on the mapping.
\textit{\textbf{Temporal reuse}} occurs when a PE accesses the same set of data across time, and can 
be leveraged via the S1 scratchpad.
%, and the data set is stored in S1 scratchpad.
For example, data 2 in~\autoref{fig:background_datareuse}  is accessed in PE0 across time 1 and 2 (red arrows).
%If PE0 has sufficient space in its local buffer (i.e., S1 scratchpad) for data 2, PE0 can reuse the data across time.
%
\textit{\textbf{Spatial reuse}} occurs when multiple PEs access the same set of data at the same time via a multicast/broadcast over the 
NoC.
%, and the underlying NoC between S2 and S1 supports multi-casting (e.g., a bus or a tree-based NoC).
For example, data 1 and 2 at time 1 in~\autoref{fig:background_datareuse} are accessed by both  PE0 and PE1 (blue arrows). 
%They can be multicast from S2 scratchpad to PE0 and PE1, reducing the number of S2 accesses.
%
%Note that spatial reuse does not refer to spatial locality; i.e., spatial reuse does not refer to data reuse within a memory block (or cache line) but multicasting of data that reduces the number of remote buffer accesses.
\textit{\textbf{Spatio-temporal reuse}} occurs when two adjacent PEs access the same set of data in a skewed schedule, and they are interconnected via point-to-point connections (i.e., wire). For example, data 2 and 3 in~\autoref{fig:background_datareuse} are accessed in PE1 at time 1 and in PE0 at time 2.
% \SR{These indices were wrong. It would have been easier and correct in description if it moved from PE0 to PE1 in figure. I fixed description. If we fix the figure fix here.}
Those data can be directly forwarded from PE1 to PE0 so that PE0 does not need to fetch the data from S2.
%Such reuse patterns can be found in systolic arrays like TPU~\cite{jouppi2017datacenter}.
The ability of an accelerator to leverage all three kinds of reuse depends on its internal microarchitecture (i.e., size of S1 and S2 buffers for temporal and NoC topology for spatial/spatio-temporal).
\emph{These reuse features have to be taken into account in addition to scratchpad sizes 
%as part of the analytical model and
when developing mapping strategies 
for any algorithm running on this hardware.}

%The specific breakdown of temporal, spatial and spatio-temporal reuse depends on the mapping, described next.
%In this work, we target spatial accelerator substrates that can support all three kinds of reuse, so that we can explore the entire space of mappings.
%We would like to point out 
%that GPUs can only exploit temporal reuse, not spatial/spatio-temporal
%reuse due to the lack of NoC support
%for multicasts and forwarding.
%Thus their mapping space is more constrained.

\betterparagraph{Comparison to GPUs}
\autoref{fig:background_mapping} also shows the differences between spatial accelerators and temporal ones such as GPUs.
Some of the major logical differences %between spatial accelerators and GPUs 
are 1) ALUs in spatial accelerators communicate directly using the NoC without register file intervention, unlike ALUs in GPU's which communicate via writing and reading from the register file, and 2) Spatial accelerators provide spatial, temporal, and spatio-temporal data reuse
%(see \autoref{subsec:reuse}), 
%i.e., \emph{spatial} reuse (same data element mapped on different PEs at the same time step) and \emph{spatio-temporal} reuse (same data element for different PEs at adjacent time steps) 
(via hardware support for multi-cast, broadcast, direct forwarding and spatial reduction as part 
of the NoC) as opposed to only temporal reuse (via scratchpads) on GPUs.
%Section ~\ref{sec:background} describes these approaches for reuse in detail.
This allows spatial accelerators to achieve higher performance, better runtime and lower energy usage.

%A  spatial accelerator~\cite{chen2016eyeriss, jouppi2017datacenter, kwon2018maeri,nvdla, du2015shidiannao} is 
%a specialized hardware 
%ASIC
%for running a particular application (e.g., DNN inference) with high energy efficiency and low latency.
%This is achieved by performing efficient HW-SW co-design, from the algorithm to compiler to hardware architecture.
%\footnote{These accelerators are sometimes called ``dataflow" accelerators as they rely on mapping the workload on the PEs and relying on parallel dataflow processing~\cite{chen2016eyeriss}, instead of sequential control-flow via a Program Counter like CPUs. However, we would like to point out that are much more restrictive than traditional dataflow processors~\cite{dennis1974preliminary,culler1986dataflow,kathail1981multiple,shimada1986evaluation,gurd1985manchester} as they target specific statically-analyzable applications such as DNNs or \gemm, simplifying the compiling problem that plagued dataflow processors.}.
%through computations stacks; from architecture to compiler and algorithm levels.
%Such dataflow patterns are determined by a mapping, which we refer to computation ordering and parallelization strategies (i.e., dataflow~\cite{chen2016eyeriss}) and data tile sizes.
%Since dataflow ultimately determines the amount and type of data reuse in accelerators, optimizing mapping is crucial in accelerator design.
%Therefore, we discuss the type of data reuse in accelerators and mapping next.

\emph{To summarize, spatial accelerators are considerably different from traditional accelerators and diverse. 
Developing a framework (Section~\ref{sec:flash}) to map widely used kernels on these accelerators and 
evaluating these accelerator designs (Section~\ref{sec:evaluation}) 
%and developing a framework (Section~\ref{sec:flash}) that is able to model these accelerators 
is a key first step in understanding their value. This is the focus of this work.}
 
 %is part of NVDLA's dataflow to optimize for modern CNNs like ResNet50~\cite{he2016deep}
 %, and so on.
% orchestrating data movement such as reuse behavior or parallelization and these are encoded into the accelerator micro-architecture (a.k.a dataflow), for example, keeping weights stationary in systolic accelerators and parallelizing input/output channels of convolutions in DLA accelerators.
%The focus of these features is better performance in ML benchmarks.

\subsection{Dataflow Directives and Mapping}
\label{subsec:background_directive_mapping}
%\SR{SR: I moved this here from previous section as it is a better fit here.}
\betterparagraph{Dataflow} Spatial accelerators carefully encode the data movement for reuse and parallelization into the accelerator's hardware micro-architecture. This is also known as ``\textbf{dataflow}''\footnote{Note that this is different from dataflow style of programming that was popular few decades ago.}.
%choice is often different 
%among different spatial accelerators depending on the microarchitecture~\cite{eyeriss_isca}.
Specifically, the dataflow includes two key components:
\textit{(i) parallelization strategy}, i.e., which dimensions of the tensors can be run in parallel.
\textit{(ii) computation order}, i.e., the order in which the dimensions of the tensor are scheduled over the accelerator.

The dataflow has two implications.
First, it determines the amount of data reuse 
on the inputs and outputs.
Prior work has studied this relationship for the ML use case ~\cite{eyeriss_isca, kwon2019understanding} and concluded that 
%no one dataflow is the best for ML use cases, 
the reuse efficiency of the dataflow depends on
 the target DNN layer dimensions and shape.
 \emph{We extend this to \gemm of different sizes and shapes by developing a framework that can find the optimal tile sizes and pruning the search space for a variety of workloads and accelerator combinations.}
Second, dataflow choices also have implications on accelerator microarchitecture aspects.
The parallelization strategy directly 
affects the NoC implementation. Also, given accelerator microarchitecture (as in this study), the dataflow choices are restricted. 
For example, 
%parallelism across inputs can lead to replication (i.e., spatial reuse) of weights in the ML use case, needing multicast support in the NoC (via a bus or tree); 
parallelism across the dimension being reduced (e.g., \textit{K} in a \gemm or input channels in a convolution) leads to different PEs computing partial sums for the same output and needs reduction via the NoC (e.g., a reduction chain or tree).
%Similarly, a computation order where the accelerator streams through the input matrix needs to keep tiles of weights \textit{stationary}~\cite{eyeriss_isca} in the scratchpad buffer, and vice versa.
\emph{We take these hardware functionality into account when mapping \gemm to an accelerator.}
Supporting complex dataflows requires more complexity in the accelerator hardware~\cite{kwon2018maeri,kwon2019understanding}.
Different spatial accelerators 
(commercial~\cite{jouppi2017datacenter, nvdla} and research prototypes~\cite{chen2016eyeriss, du2015shidiannao}) 
have picked 
different dataflows~\cite{eyeriss_isca} 
trading off reuse-efficiency and hardware complexity.

% The relationship between dataflows and data reuse 
% has been studied in prior work~\cite{eyeriss_isca, kwon2019understanding} and 
% concluded that no one dataflow is the best; 

%different dataflows works 
%better depending on the dimensions 
%of convolutional 
%layers (that most of these spatial accelerators target).
\insertWideFigure{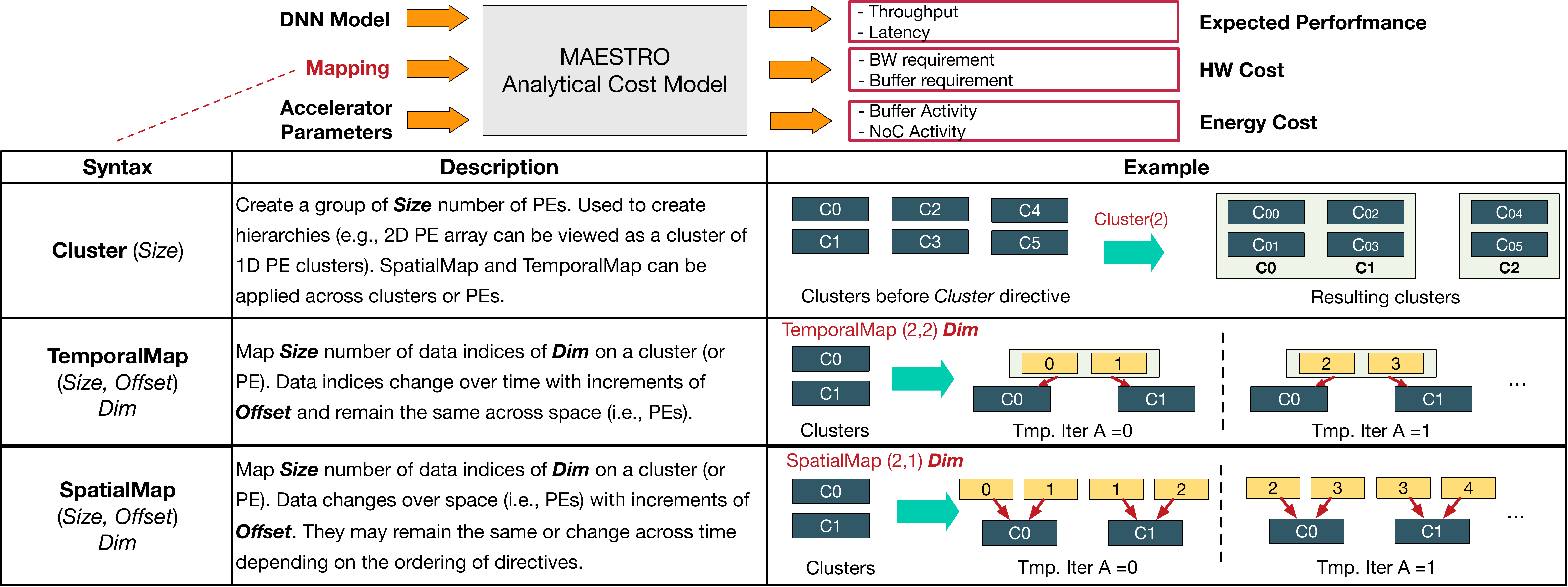}{Syntax and description of MAESTRO dataflow directives.}
%\insertFigure{maestro_directives}{\rev{Syntax and Description of MAESTRO dataflow directives.}}
\betterparagraph{Dataflow Directives}
In this work, we leverage the \textit{dataflow directives} introduced by MAESTRO~\cite{kwon2019understanding} to express the exact dataflows for various accelerators under consideration.

There are three directives: {\sf TemporalMap}, {\sf SpatialMap}, and {\sf Cluster} which are described in \autoref{fig:maestro_directives}.
%We summarize the key intuition:
{\sf TemporalMap} implies that the data changes over time, and remains same over space (i.e., across PEs). %, 
%offering opportunities for spatial and spatio-temporal reuse.
SpatialMap implies that the data changes over space (i.e., parallelism). %, and offers opportunity for temporal reuse.
{\sf Cluster} helps describe hierarchical dataflows by grouping PEs into clusters of certain {\sf Size}. 

One could recursively define dataflows within clusters. For example, NVDLA~\cite{nvdla} maps convolutions by employing {\sf SpatialMap} across input channels as its intra-cluster dataflow, and 
{\sf SpatialMap} across output channels
as its inter-cluster dataflow. 
The relative order among 
the directives 
specifies computation order.
We provide a concrete example later in \autoref{subsec:walk_through_example}.
The specific amount 
of reuse depends 
on the overall dataflow, 
workload dimensions 
and tile sizes (together called mapping).

\begin{table}[ht]
\centering
\caption{Specifications of target spatial accelerator architectures. Note that \textit{input-}, \textit{weight-}, and \textit{output-stationary} dataflows correspond to input matrix \textit{A}-, input matrix \textit{B}-, and output matrix \textit{C}-stationary dataflows in \gemm operation, respectively. More details of the dataflow are in \autoref{table:flash_mapping_comparison}. For our evaluations, we provide equal hardware resources (PEs and buffers) to all accelerators (\autoref{table:hardware_config}).}
\scalebox{0.82}{
\begin{tabular}{c|c|c|c}
\thickhline
\multirow{2}{*}{\begin{tabular}[c]{@{}c@{}}\textbf{Spatial}\\ \textbf{Accelerators}\end{tabular}} & \multicolumn{2}{c|}{\textbf{HW Configuration}}                            & \multirow{2}{*}{\begin{tabular}[c]{@{}c@{}}\textbf{Dataflow}\end{tabular}} \\ \cline{2-3}
                                                                                & \textbf{\# PEs (row$\times$col)}  & \begin{tabular}[c]{@{}c@{}}\textbf{NoC}\end{tabular} &                                                                                       \\ \hline
Eyeriss~\cite{chen2016eyeriss}                                                                         & 12$\times$14 & Buses                                                     & \textit{input row stationary}                                                                  \\ \hline
NVDLA~\cite{nvdla}                                                                           & 64$\times$8 &  Bus + Tree                                                     & \textit{weight-stationary}                                                                     \\ \hline
TPUv2~\cite{jouppi2017datacenter}                                                                             & 128$\times$128 & Mesh                                                     & \textit{weight-stationary}                                                                     \\ \hline
ShiDianNao~\cite{du2015shidiannao}                                                                      & 8$\times$8 & Mesh                                                     & \textit{output-stationary}                                                                     \\ \hline
MAERI~\cite{kwon2018maeri}                                                                           & \begin{tabular}[c]{@{}c@{}}256\\(any aspect ratio)\end{tabular} & \begin{tabular}[c]{@{}c@{}}Custom\\Fat Tree\end{tabular}                            & \textit{flexible}                                                                                \\ \thickhline
\end{tabular}
}
\label{table:target_accelerators}
\end{table}
\def\arraystretch{1.2}%
\begin{table*}[ht]
\centering
\caption{Comparison of various \gemm mappings constrained by state-of-the-art spatial accelerators where \textit{P} and $\lambda$ denote total number of PEs and size of cluster (i.e., number of PEs in each cluster), respectively. $\textit{T}^{\textit{out}/\textit{in}}_{\textit{dim.}}$ denotes the tile size for each of dimensions \textit{M}, \textit{N} and \textit{K} in the outer/inner clusters. {\sf TMap} and {\sf SMap} refer to {\sf TemporalMap} and {\sf SpatialMap} directives, respectively.}
\scalebox{0.71}{
\begin{tabular}{c||c|c|c|c|c}
\thickhline
\textbf{\begin{tabular}[c]{@{}c@{}}\large{Spatial Accelerator}\end{tabular}} &  \large{Eyeriss}~\cite{chen2016eyeriss} & \large{NVDLA}~\cite{nvdla} & \large{TPU}~\cite{jouppi2017datacenter} & \large{ShiDianNao}~\cite{du2015shidiannao} & \large{MAERI}~\cite{kwon2018maeri}\\ \hline \hline

\textbf{\begin{tabular}[c]{@{}c@{}}\large{Dataflow:}\\\large{Parallel Dim}\end{tabular}}
& 

\begin{tabular}[c]{@{}c@{}}Inter-Cluster: \textit{M}\\Intra-Cluster: \textit{K} \end{tabular} &

\begin{tabular}[c]{@{}c@{}}Inter-Cluster: \textit{N}\\Intra-Cluster: \textit{K}\end{tabular}& 

\begin{tabular}[c]{@{}c@{}}Inter-Cluster: \textit{N}\\Intra-Cluster: \textit{K}\end{tabular} & 

\begin{tabular}[c]{@{}c@{}}Inter-Cluster: \textit{M}\\Intra-Cluster: \textit{N}\end{tabular} & 
\begin{tabular}[c]{@{}c@{}}Inter-Cluster: \textit{M} or \textit{N} or \textit{K}\\Intra-Cluster: \textit{M} or \textit{N} or \textit{K}\end{tabular}\\ \hline

\textbf{\begin{tabular}[c]{@{}c@{}}\large{Dataflow:}\\\large{Compute Order}\end{tabular}}  & 

\begin{tabular}[c]{@{}c@{}}Inter-Cluster: \textless{}\textit{m}, \textit{n}, \textit{k}\textgreater{}\\Intra-Cluster: \textless{}\textit{m}, \textit{n}, \textit{k}\textgreater{}\end{tabular} &

\begin{tabular}[c]{@{}c@{}}Inter-Cluster: \textless{}\textit{n}, \textit{k}, \textit{m}\textgreater{}\\Intra-Cluster: \textless{}\textit{n}, \textit{m}, \textit{k}\textgreater{}\end{tabular} & 

\begin{tabular}[c]{@{}c@{}}Inter-Cluster: \textless{}\textit{n}, \textit{m}, \textit{k}\textgreater{}\\Intra-Cluster: \textless{}\textit{n}, \textit{m}, \textit{k}\textgreater{}\end{tabular} & 

\begin{tabular}[c]{@{}c@{}}Inter-Cluster: \textless{}\textit{m}, \textit{n}, \textit{k}\textgreater{}\\Intra-Cluster: \textless{}\textit{m}, \textit{n}, \textit{k}\textgreater{}\end{tabular} &

\begin{tabular}[c]{@{}c@{}}Inter-Cl.: \textless{}\textit{m/n/k}, \textit{n/m/k}, \textit{k/m/n}\textgreater{} \\Intra-Cl.: \textless{}\textit{m/n/k}, \textit{n/m/k}, \textit{k/m/n}\textgreater{} \end{tabular}\\ \hline

\textbf{\large{Cluster Size ($\lambda$)}}  & 

\begin{tabular}[c]{@{}c@{}}1 $\leq$ $\lambda$ $\leq$ 12\\(compile time-flexible)\end{tabular} & 

\begin{tabular}[c]{@{}c@{}}16 $\leq$ $\lambda$ $\leq$ 64\\(design time-flexible)\end{tabular} & 

256 or $\sqrt{\textit{P}}$ & 

8 or $\sqrt{\textit{P}}$ & 
\begin{tabular}[c]{@{}c@{}}$\textit{T}^{out}_{\textit{M}}$\textit{/}$\textit{T}^{out}_{\textit{N}}$\textit{/}$\textit{T}^{out}_{\textit{K}}$\\(tile size of the last dimension)\end{tabular}\\ \hline

\textbf{\begin{tabular}[c]{@{}c@{}}\large{\gemm Mapping}\end{tabular}}                                                 &

\multicolumn{1}{l|}{\begin{tabular}[c]{@{}l@{}}\textcolor{violet}{{\sf \textbf{SMap}}} ($\textit{T}^{out}_{\textit{M}}$,$\textit{T}^{out}_{\textit{M}}$) \textit{M}\\ \textcolor{violet}{{\sf \textbf{TMap}}} ($\textit{T}^{out}_{\textit{N}}$,$\textit{T}^{out}_{\textit{N}}$) \textit{N}\\ \textcolor{violet}{{\sf \textbf{TMap}}} ($\textit{T}^{out}_{\textit{K}}\times$$\lambda$,$\textit{T}^{out}_{\textit{K}}\times$$\lambda$) \textit{K}\\ \textcolor{black}{{\sf \textbf{Cluster}}} ($\lambda$)\\ \textcolor{violet}{{\sf \textbf{TMap}}} ($\textit{T}^{in}_{\textit{M}}$,$\textit{T}^{in}_{\textit{M}}$) \textit{M}\\ \textcolor{violet}{{\sf \textbf{TMap}}} ($\textit{T}^{in}_{\textit{N}}$,$\textit{T}^{in}_{\textit{N}}$) \textit{N}\\ \textcolor{violet}{{\sf \textbf{SMap}}} ($\textit{T}^{out}_{\textit{K}}$,$\textit{T}^{out}_{\textit{K}}$) \textit{K}\end{tabular}} &

\multicolumn{1}{l|}{\begin{tabular}[c]{@{}l@{}}\textcolor{violet}{{\sf \textbf{SMap}}} ($\textit{T}^{out}_{\textit{N}}$,$\textit{T}^{out}_{\textit{N}}$) \textit{N}\\ \textcolor{violet}{{\sf \textbf{TMap}}} ($\textit{T}^{out}_{\textit{K}}\times$$\lambda$,$\textit{T}^{out}_{\textit{K}}\times$$\lambda$) \textit{K} \\ \textcolor{violet}{{\sf \textbf{TMap}}} ($\textit{T}^{out}_{\textit{M}}$,$\textit{T}^{out}_{\textit{M}}$) \textit{M}\\ \textcolor{black}{{\sf \textbf{Cluster}}} ($\lambda$)\\ \textcolor{violet}{{\sf \textbf{TMap}}} ($\textit{T}^{in}_{\textit{N}}$,$\textit{T}^{in}_{\textit{N}}$) \textit{N}\\ \textcolor{violet}{{\sf \textbf{TMap}}} ($\textit{T}^{in}_{\textit{M}}$,$\textit{T}^{in}_{\textit{M}}$) \textit{M}\\ \textcolor{violet}{{\sf \textbf{SMap}}} ($\textit{T}^{out}_{\textit{K}}$,$\textit{T}^{out}_{\textit{K}}$) \textit{K}\end{tabular}}  &

\multicolumn{1}{l|}{\begin{tabular}[c]{@{}l@{}}\textcolor{violet}{{\sf \textbf{SMap}}} ($\textit{T}^{out}_{\textit{N}}$,$\textit{T}^{out}_{\textit{N}}$) \textit{N}\\ \textcolor{violet}{{\sf \textbf{TMap}}} ($\textit{T}^{out}_{\textit{M}}$,$\textit{T}^{out}_{\textit{M}}$) \textit{M}\\ \textcolor{violet}{{\sf \textbf{TMap}}} ($\textit{T}^{out}_{\textit{K}}\times$$\lambda$,$\textit{T}^{out}_{\textit{K}}\times$$\lambda$) \textit{K}\\ \textcolor{black}{{\sf \textbf{Cluster}}} ($\lambda$)\\ \textcolor{violet}{{\sf \textbf{TMap}}} ($\textit{T}^{in}_{\textit{N}}$,$\textit{T}^{in}_{\textit{N}}$) \textit{N}\\ \textcolor{violet}{{\sf \textbf{TMap}}} ($\textit{T}^{in}_{\textit{M}}$,$\textit{T}^{in}_{\textit{M}}$) \textit{M}\\ \textcolor{violet}{{\sf \textbf{SMap}}} ($\textit{T}^{out}_{\textit{K}}$,$\textit{T}^{out}_{\textit{K}}$) \textit{K}\end{tabular}}  &

\multicolumn{1}{l|}{\begin{tabular}[c]{@{}l@{}}\textcolor{violet}{{\sf \textbf{SMap}}} ($\textit{T}^{out}_{\textit{M}}$,$\textit{T}^{out}_{\textit{M}}$) \textit{M}\\\textcolor{violet}{{\sf \textbf{TMap}}} ($\textit{T}^{out}_{\textit{N}}\times$$\lambda$,$\textit{T}^{out}_{\textit{N}}\times$$\lambda$) \textit{N}\\ \textcolor{violet}{{\sf \textbf{TMap}}} ($\textit{T}^{out}_{\textit{K}}$,$\textit{T}^{out}_{\textit{K}}$) \textit{K}\\ \textcolor{black}{{\sf \textbf{Cluster}}} ($\lambda$)\\\textcolor{violet}{{\sf \textbf{TMap}}} ($\textit{T}^{in}_{\textit{M}}$,$\textit{T}^{in}_{\textit{M}}$) \textit{M}\\ \textcolor{violet}{{\sf \textbf{SMap}}} ($\textit{T}^{out}_{\textit{N}}$,$\textit{T}^{out}_{\textit{N}}$) \textit{N}\\\textcolor{violet}{{\sf \textbf{TMap}}} ($\textit{T}^{in}_{\textit{K}}$,$\textit{T}^{in}_{\textit{K}}$) \textit{K}\end{tabular}} &

\multicolumn{1}{l}{\begin{tabular}[c]{@{}l@{}}\textcolor{violet}{{\sf \textbf{TMap}}} ($\textit{T}^{out}_{\textit{M}}$,$\textit{T}^{out}_{\textit{M}}$) \textit{M}\\ \textcolor{violet}{{\sf \textbf{SMap}}} ($\textit{T}^{out}_{\textit{N}}$,$\textit{T}^{out}_{\textit{N}}$) \textit{N}\\ \textcolor{violet}{{\sf \textbf{TMap}}} ($\textit{T}^{out}_{\textit{K}}$,$\textit{T}^{out}_{\textit{K}}$) \textit{K}\\ \textcolor{black}{{\sf \textbf{Cluster}}} ($\textit{T}^{out}_{\textit{K}}$)\\ \textcolor{violet}{{\sf \textbf{TMap}}} ($\textit{T}^{in}_{\textit{M}}$,$\textit{T}^{in}_{\textit{M}}$) \textit{M}\\ \textcolor{violet}{{\sf \textbf{TMap}}} ($\textit{T}^{in}_{\textit{N}}$,$\textit{T}^{in}_{\textit{N}}$) \textit{N}\\ \textcolor{violet}{{\sf \textbf{SMap}}} (1,1) \textit{K}\end{tabular}}\\ \hline

\textbf{\begin{tabular}[c]{@{}c@{}}\large{Mapping Name}\end{tabular}}    & \begin{tabular}[c]{@{}c@{}}{{{\sf STT\_TTS-MNK}}} \\ {(Eyeriss-style)} \end{tabular} & \begin{tabular}[c]{@{}c@{}}{{{\sf STT\_TTS-NKM}}} \\{(NVDLA-style)} \end{tabular} & \begin{tabular}[c]{@{}c@{}}{{{\sf STT\_TTS-NMK}}} \\{(TPU-style)} \end{tabular} & \begin{tabular}[c]{@{}c@{}}{{{\sf STT\_TST-MNK}}} \\{(ShiDianNao-style)} \end{tabular} &  \begin{tabular}[c]{@{}c@{}}{{{\sf TST\_TTS-MNK}}} \\{(MAERI-style)} \end{tabular}\\ \thickhline

% \textbf{\begin{tabular}[c]{@{}c@{}}\large{Feasible}\\\large{Loop Orders}\end{tabular}} &

% \begin{tabular}[c]{@{}c@{}}\textit{\textless{}m, n, k\textgreater{}}, \textit{\textless{}n, m, k\textgreater{}},\\\textit{\textless{}m, k, n\textgreater{}}, \textit{\textless{}n, k, m\textgreater{}},\\\textit{\textless{}k, m, n\textgreater{}}, \textit{\textless{}k, n, m\textgreater{}}\end{tabular} &

% \begin{tabular}[c]{@{}c@{}}\textit{\textless{}m, n, k\textgreater{}}, \textit{\textless{}n, m, k\textgreater{}},\\\textit{\textless{}m, k, n\textgreater{}}, \textit{\textless{}n, k, m\textgreater{}}\end{tabular} &

% \begin{tabular}[c]{@{}c@{}}\textit{\textless{}m, n, k\textgreater{}}, \textit{\textless{}n, m, k\textgreater{}},\\\textit{\textless{}m, k, n\textgreater{}}, \textit{\textless{}n, k, m\textgreater{}}\end{tabular} &

% \begin{tabular}[c]{@{}c@{}}\textit{\textless{}m, n, k\textgreater{}}, \textit{\textless{}n, m, k\textgreater{}},\\\textit{\textless{}m, k, n\textgreater{}}, \textit{\textless{}n, k, m\textgreater{}}\end{tabular} &

% \begin{tabular}[c]{@{}c@{}}\textit{\textless{}m, n, k\textgreater{}}, \textit{\textless{}n, m, k\textgreater{}},\\\textit{\textless{}m, k, n\textgreater{}}, \textit{\textless{}n, k, m\textgreater{}}\end{tabular} \\ \hline

\end{tabular}
}
\label{table:flash_mapping_comparison}
\end{table*}

\betterparagraph{Mapping} 
The number of compute units and amount of on-chip buffers in accelerators are typically not sufficient to cover entire workload (e.g., \blas operations on a node are on matrices of size O($10^3-10^4$) and Resnet50~\cite{he2016deep} contains $3.8 \times 10^{9}$ FLOPs while accelerators have hundreds~\cite{ieee2016eyeriss} to thousands~\cite{jouppi2017datacenter} PEs).
Therefore, tiling the computation and data of workload and time-multiplexing the target accelerator is necessary.

The relationship between mapping and dataflow is shown in ~\autoref{fig:background_mapping}.
Specifically, 
the dataflow of the accelerator, 
the tile sizes for all tensors, 
and  
scheduling of these tiles 
for the entire workload
is known as a \textbf{mapping},
as shown in \autoref{fig:background_mapping}.
The mapping determines the data needed 
at each PE at each instance of time.
The proportion of this data that needs to be moved across the memory hierarchy (from off-chip DRAM to S2, and from S2 to S1) and the proportion that can 
be reused depends on the accelerator's dataflow.
%It can be used to infer data reuse across time and space, which in-turn can be exploited by an accelerator if it has proper hardware support for each type of data reuse, as described above.
%Thus, it determines the total data-movement (from off-chip DRAM to S2, and from S2 to S1) for running the workload.

An optimized mapping is crucial for overall efficiency, since data movement dominates energy in accelerators~\cite{ieee2016eyeriss}.
The {\sf Size} parameter in the \maestro's {\sf SpatialMap} and {\sf TemporalMap} directives in \autoref{fig:maestro_directives} can be used to specify the tile sizes for
each dimension of the matrices.
\section{Accelerator Modeling Methodology}
\label{sec:target_accelerators}

%\subsection{Accelerators in this study}
%\label{subsec:accelerators_this_study}
% \rev{We briefly highlight the key features of the accelerators that are part of our study here and refer the reader to other sources for more details~\cite{}.}

% \input{tables/flash_all_mappings}
\autoref{table:target_accelerators} shows the key features of five target spatial accelerators explored in our study. The diversity of the chosen accelerators adds to the complexity of evaluating even one kernel on all of them. We discuss our methodology below.
%The analytical model, tiling strategies, and pruning strategies all have to be adapted to these.

% SR This is in the table now.
%The 128$\times$128 TPU dataflow keeps the \textit{weight-stationary} at the PEs so input vectors can be streamed in while the parallelism is over the \emph{output channel}. 
%% optimizing for reusing the model across multiple batches of inputs. 
%The 512 MAC NVDLA is also \textit{weight-stationary}, but it parallelizes the input/output channels of convolutions.
%Eyeriss keeps \textit{input rows-stationary} and allows reductions that are flexible whereas the reductions in both TPU and NVDLA are not. ShiDianNao uses \textit{output-stationary} approach which uses a tiled approach for output activation. MAERI is the most flexible among the accelerators we study. It allows arbitrary length reductions while accumulating on the same dimension as Eyeriss. These choices to address an ML use case lead to different constraints in these hardware such as which dimension can be reduced (see Section~\ref{sec:flash}).
% \input{tables/flash_all_mappings}
\subsection{Modeling Spatial Accelerators using Dataflow Directives}
\label{subsec:modeling_spatial_accelerators}
%\TODO{Tushar has the lock. Rename mapping directives to dataflow directives and make sure section is coherent}
A key challenge in performing an apples-to-apples comparison of the accelerators described above is the fact that they differ in the number of PEs, buffer sizes, dataflow, internal microarchitecture (and software stacks~\cite{jouppi2017datacenter,nvdla} or lack-there-of for research prototypes~\cite{chen2016eyeriss, kwon2018maeri,du2015shidiannao}). 
To address this issue, 
we contrast the accelerators based on ``how" they map \gemm on the spatial substrate\footnote{Note that except for the TPU, the other four accelerators are originally designed to directly run convolutions and leverage reuse across sliding windows. We map GEMM on these 
convolution accelerators 
by expressing it as a convolution 
with one row and one channel.
%, though they could run \gemm via Im2Col and use our proposed mappings.
To the best of our knowledge, previous studies at mapping GEMM have not addressed how to map GEMM efficiently on existing state-of-the-art spatial accelerators.}. 
%Recall that a mapping comprises of the accelerator's internal dataflow (i.e., supported compute order  and parallelization, constrained by the internal microarchitecture and connectivity) and tile sizes (constrained by scratchpad sizes).
Furthermore, in our evaluations, 
we provide the same hardware resources to all accelerators (\autoref{table:hardware_config})\footnote{This is why we refer to our mapping descriptions as *-style as they are not running on the actual accelerator instances.}.

% \insertWideFigure{gemm_mapping_example}{Walk-through Example of {\sf TST\_TTS-MNK} (i.e., MAERI-style) \gemm mapping. (a) shows the example GEMM with for loops and (b) shows the abstract picture of the target accelerator in this example. (c) is the example mapping that is described using the aforementioned dataflow directives. (d) shows the assignment of entries of input matrices \textit{A} and \textit{B} to the PEs in four clusters. Each cluster computes one entry of output matrix \textit{C}. (e) shows the same from the matrix view, which PEs have to access which entries of matrices \textit{A} and \textit{B} and which PEs contribute to an entry of matrix \textit{C}.}

\insertWideFigure{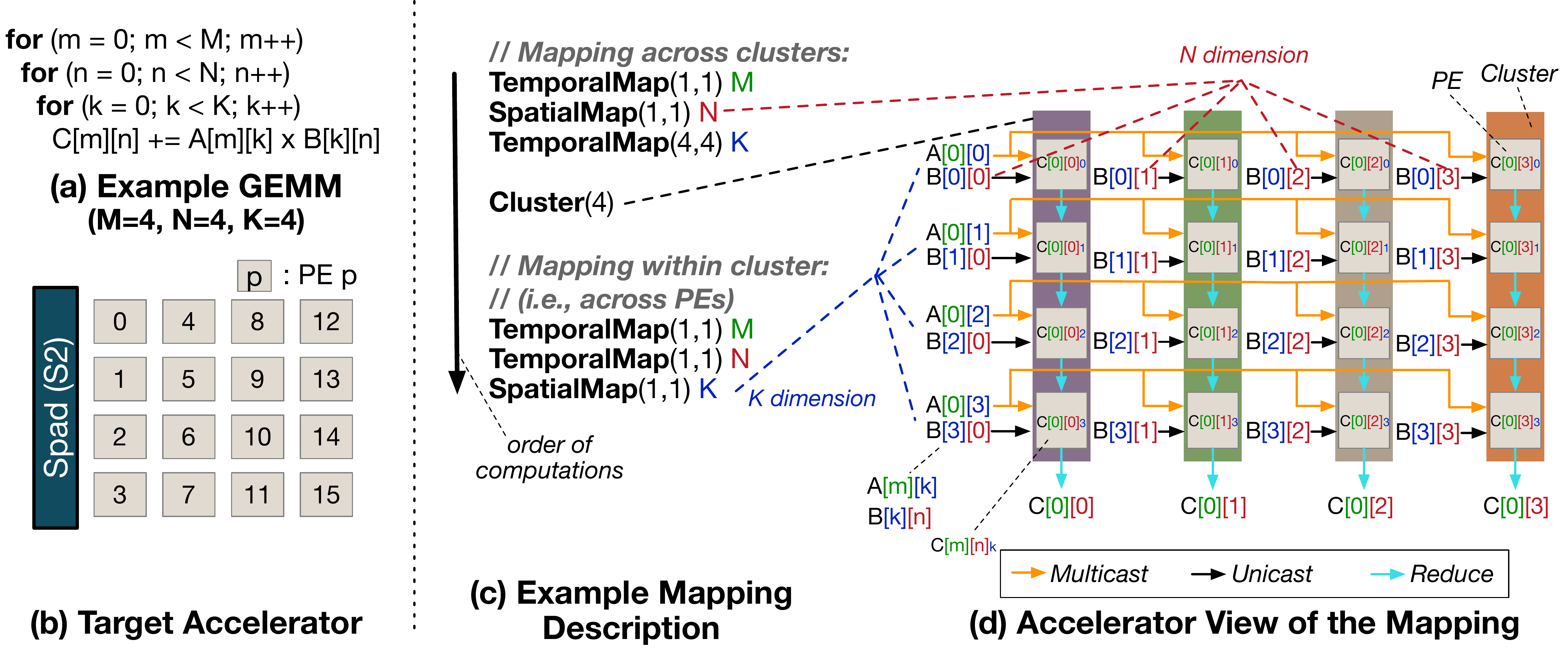}{Walk-through example of {\sf TST\_TTS-MNK} (i.e., MAERI-style) \gemm mapping. (a) shows the example GEMM with for loops and (b) shows the abstract picture of the target accelerator in this example. (c) is the example mapping that is described using the aforementioned dataflow directives. (d) shows the assignment of entries of input matrices \textit{A} and \textit{B} to the PEs in four clusters. Each cluster computes one entry of output matrix \textit{C}. We evaluate five such accelerators with different hardware configurations.}

\betterparagraph{Example of \gemm Mapping}
We employ the dataflow directives described in \autoref{subsec:background_directive_mapping} to analyze and describe how each accelerator runs a \gemm, 
constrained by their dataflow.
We refer to the \gemm mappings using the abbreviated directive order, where we use {\sf \textbf{T}} for {\sf \textbf{T}emporalMap}, {\sf \textbf{S}} for {\sf \textbf{S}patialMap} and {\sf \textbf{\_}} for introducing a cluster (i.e., hierarchy).
For example, a simple {\sf \textbf{TTT}} mapping is nothing but triple nested \gemm loop on sequential core with the loop order determined by the binding of \textit{M}, \textit{N} and \textit{K} loops (\autoref{subsec:background_gemm}) to each directive; {\sf \textbf{TTT\_TTT}} is the hierarchical/cache-blocked version of the same. An {\sf \textbf{STT}} mapping with {\sf MNK} as the loop order would be a triple nested \gemm where rows of matrix \textit{A} (i.e., dimension $M$) are spatially mapped (i.e., parallelized) across CPU cores or GPU CUDA threads or PEs in a spatial accelerator.

\betterparagraph{\gemm Mappings supported by Spatial Accelerators}
\autoref{table:flash_mapping_comparison} presents the \gemm mappings 
that can be supported by our target accelerators.
Each mapping has directives for describing the dataflows both across the clusters (written above the {\sf Cluster} directive) and within the cluster (written below the {\sf Cluster} directive).
\emph{This implies that all 
\gemm mappings use a 2D-tiled approach}.
The cluster size is tied to accelerator microarchitecture (e.g., it represents the number of PE rows in \autoref{table:target_accelerators}), while the maximum tile size is constrained by S1 and S2 sizes.

Each mapping encodes the  dataflow and hardware 
constraints from the specific accelerator.
We provide some details next.
As mentioned in \autoref{subsec:background_directive_mapping}, 
the dataflow determines 
the \emph{parallelizing dimension}, and \emph{compute order}.

We first focus on the parallelizing dimension.
Recall that the $K$ dimension 
in a \gemm is reduced when generating the output.
Eyeriss, TPU, NVDLA and MAERI all provide NoC support for spatial reduction; Eyeriss and TPU via store-and-forward across the column 
and NVDLA and MAERI via reduction trees. 
Thus, for \gemm, we map the $K$ dimension spatially within
each cluster for these four accelerators. 
Outside the cluster, the same dimension needs to be mapped temporally, with a tile size that can cover the tiles needed by all the clusters. Either $M$ or $N$ can be mapped spatially outside the clusters.
In ShiDianNao, however, there is no support for spatial reduction, and hence the $K$ dimension is mapped temporally; the parallelism comes from $N$ dimension instead within the cluster.

Next, the compute (or loop order) is determined by the
relative order of the temporal directives.
The TPU and NVDLA keep the weight matrix (i.e., matrix $B$) in \gemm stationary, and stream the input matrix (i.e., matrix $A$). This is reflected by keeping the $N$ dimension as the outermost loop, both within and across the clusters.
Eyeriss keeps the input rows stationary, and this is 
specified by keeping $M$ as the outermost dimension.
ShiDianNao keeps the output stationary, forcing a 
$M$ followed by $N$ loop order.
MAERI allows all loop orders\footnote{Accelerators like LAP~\cite{pedram2012codesign} 
implement \gemm with SUMMA algorithm.
SUMMA is a subset of the MAERI-style {\sf TST\_TTS} mapping with the \textless{}\textit{k}, \textit{m}, \textit{n}\textgreater{}/\textless{}\textit{k}, \textit{n}, \textit{m}\textgreater{} loop order.}.
In each mapping, the tile sizes are free variables, constrained by the S1 buffer (for intra-cluster) and S2 buffer (for inter-cluster).
Our policy for determining optimized tile sizes for \gemm mappings is described in \autoref{sec:flash}.

%(refer to the third column in Table \ref{table:tst_tts_mapping}).
%This implies LAP-style~\cite{pedram2012codesign} mapping is also considered part of {\sf TST\_TTS} mapping.

%%%%%%%%%%%%%%%%%%%%%%%%
% SR: Already covered above
%Note that, if we apply interchangeability rules to \textit{STT\_TTS} mapping, it is possible that three different mappings motivated by Eyeriss, NVDLA and TPU can have exactly the same loop order. However, those mappings can still be distinguished by their different cluster size.

%In the next sub-section, we demonstrate details of \textit{TST\_TTS} mapping motivated by MAERI which is the only mapping scheme that can support all loop order algorithms of \gemm based on its flexible hardware support.

%\subsection{Walk-through Example of \gemm mapping}

\subsection{Walk-through Example of {\sf \textbf{TST\_TTS}} Mapping}
\label{subsec:walk_through_example}
%The \gemm operation is the backbone of many \blas operations, and thus, supporting \gemm operation is our key extension to \maestro. As highlighted in \autoref{fig:FLASH_Overview}(a), 
%We extended \maestro's input format to accept \gemm as another type of input layer.
%Second, we enhance the \textbf{\textit{backend analytical cost model}} to compute data reuse and corresponding performance, which we describe later in this section.

%\betterparagraph{Walk-through Example of a \gemm mapping}
\autoref{fig:gemm_mapping_example}(a) shows a simple \gemm (with \textit{M}=4, \textit{N}=4 and \textit{K}=4) which we want to map on a 
16 PE spatial accelerator (\autoref{fig:gemm_mapping_example}(b)).
\autoref{fig:gemm_mapping_example}(c) shows an example mapping for this \gemm using the dataflow directives.
%discussed above.
Note that this {\sf TST\_TTS} (MAERI-style) mapping is one of many possible mappings as we discussed in Section~\ref{subsec:modeling_spatial_accelerators}.
% We describe three key facets of this mapping, that can be seen in 
% \autoref{fig:gemm_mapping_example}(d) and \autoref{fig:gemm_mapping_example}(e) from the perspective of the accelerator and the workload respectively: Clustering, Intra-Cluster Behavior, and Inter-Cluster Behavior.
We describe three key facets of this mapping, that can be seen in 
\autoref{fig:gemm_mapping_example}(d) from the perspective of the accelerator: Clustering, Intra-Cluster Behavior, and Inter-Cluster Behavior.

(i) \textit{Clustering.} In this mapping, the 16 PEs are divided into four clusters, 4 PEs each (specified via the {\sf Cluster} directive) corresponding to the accelerator columns in \autoref{fig:gemm_mapping_example}(d). This allows the mapping to exploit the 2D PE array by spatially distributing two GEMM dimensions.

(ii) \textit{Intra-Cluster Behavior.} Directives below the cluster directive specify the intra-cluster mapping. Within each cluster (i.e., column of the accelerator in \autoref{fig:gemm_mapping_example}(d)), the \textit{K} dimension is mapped spatially (specified via the {\sf SpatialMap}), while the \textit{M} and \textit{N} dimensions are mapped temporally (i.e., remain same across all the PEs and only change with time). All directives use a \textit{Size} and \textit{Offset} of 1.
These directives specify that each PE receives \textit{one} unique element from the row and column of the matrices \textit{A} and \textit{B} respectively, since \textit{M} and \textit{N} stay the same, but \textit{K} changes in each PE. This can be visualized from \autoref{fig:gemm_mapping_example}(d).
During this operation, each PE computes a partial sum and forwards it to its neighbor along the column for accumulation. Each cluster thus computes one element of matrix $C$. % via a dot-product.

(iii) \textit{Inter-Cluster Behavior.} Directives above the cluster directive specify the inter-cluster mapping. Across the clusters, the \textit{N} dimension is mapped spatially, while the \textit{M} and \textit{K} dimensions are temporal.
This means that the elements of the matrix \textit{B} 
%(i.e., \textit{K}$\times$\textit{N}) 
gets distributed across the different clusters, while the elements of the matrix \textit{A} 
%(i.e., \textit{M}$\times$\textit{K})
remain the same (and can thus be multicast).
The \textit{size} and \textit{offset} field for \textit{K} is 4 to specify that each cluster receives 4 elements from each matrix (which then get distributed among the 4 PEs within the cluster as discussed above). If this field is not set appropriately, it can lead to under-utilization within the cluster. \flash takes care of this as we discuss later in \autoref{sec:flash}.
% From \autoref{fig:gemm_mapping_example}(e), it can be seen that each time-step of the mapping computes one row of outputs for the matrix \textit{C}, and would move on to the next row in the next time-step.
From \autoref{fig:gemm_mapping_example}(d), it can be seen that each time-step of the mapping computes one row of outputs $C_{0,:}$ for the matrix \textit{C}, and would move on to the next row in the next time-step.
If the dimensions of the matrix exceed the dimensions of the physical array, the computation would need to be tiled.
The computation order across tiles of the three matrices depends on the order in which the directives are specified (\autoref{fig:gemm_mapping_example}(c)).%\GM{@Tushar: Fig. 5(b) shows a 2D PE array. I think this should point to Fig. 5(c) or Fig. 5(d)}

\betterparagraph{Example of Tiling in MAERI-style ({\sf \textbf{TST\_TTS}}) Mapping}
% \label{subsubsec:TST_TTS_example}
% \input{tables/flash_TST_TTS_mapping}
%
%% TK : removed to fit for PPoPP new template
% Table \ref{table:tst_tts_mapping} lists MAERI-style mappings for three different loop orders. The order of the dimensions \textit{M}, \textit{N} and \textit{K} for outer/inner clusters in {\sf TST\_TTS} is consistent with the loop order in the original \gemm program. For example, \textless{}\textit{i}, \textit{j}, \textit{k}\textgreater{} loop order is given by \textit{M}, \textit{N}, \textit{K} dimension order for inner/outer clusters. Using appropriate tile sizes is critical for performance of any mapping, as we show later in \autoref{subsec:optimal_tile_size}. We denote $\textit{T}^{out}_{\textit{M}}$ and $\textit{T}^{in}_{\textit{M}}$ as the \textbf{tile sizes} of \textit{M} dimension in the outer cluster and inner cluster, respectively.
%
\insertFigure{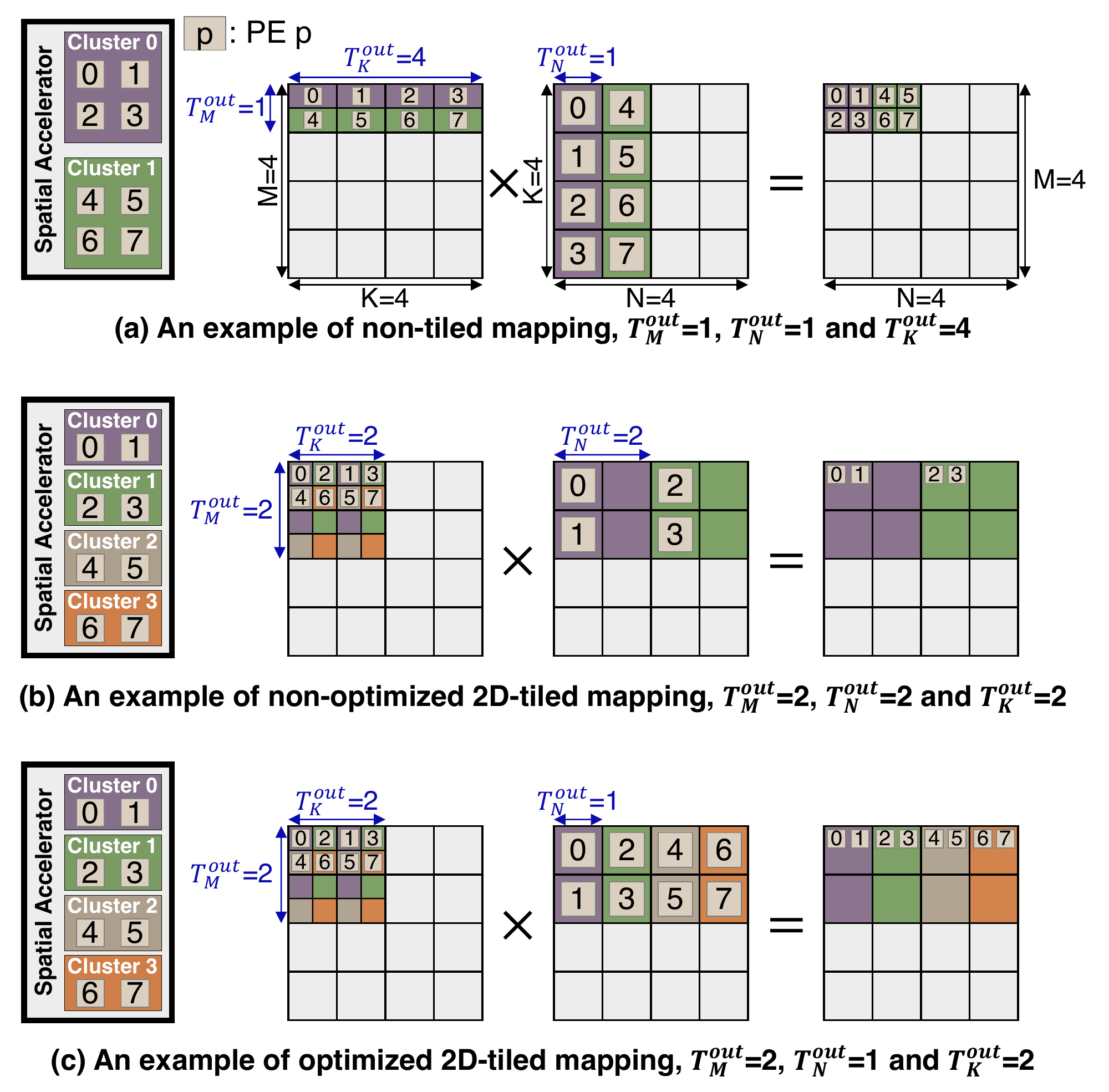}{Three examples of {\sf TST\_TTS-MNK} mapping when the tile sizes $\textit{T}^{out}_{\textit{M}}$, $\textit{T}^{out}_{\textit{N}}$ and $\textit{T}^{out}_{\textit{K}}$ are varied. For all examples, \textit{M}=4, \textit{N}=4, \textit{K}=4, total number of PEs \textit{P}=8, $\textit{T}^{in}_{\textit{M}}$=$\textit{T}^{out}_{\textit{M}}$, $\textit{T}^{in}_{\textit{N}}$=$\textit{T}^{out}_{\textit{N}}$ and $\textit{T}^{in}_{\textit{K}}$=1.}
\autoref{fig:modeling_FLASH_GEMM_example} shows the impact of different tile sizes on the matrices \textit{A} and \textit{B} of size 4$\times$4 with eight PEs. We use the notation $\textit{T}^{out}_{\textit{M}}$ (${T}^{in}_{\textit{N}}$) for tile size of \textit{M} (\textit{N}) dimension in outer (inner) loop or  inter-cluster (intra-cluster).  
Assume that $\textit{T}^{out}_{\textit{M}}$=$\textit{T}^{in}_{\textit{M}}$=1, $\textit{T}^{out}_{\textit{N}}$=$\textit{T}^{in}_{\textit{N}}$=1, $\textit{T}^{out}_{\textit{K}}$=4, and $\textit{T}^{in}_{\textit{K}}$=1 as shown in~\autoref{fig:modeling_FLASH_GEMM_example} (a). 
%In the {\sf TST\_TTS} mapping for \textless{}\textit{m}, \textit{n}, \textit{k}\textgreater{} loop order (shown in the first column of Table \ref{table:flash_mapping_comparison}), the cluster size is chosen to be same as the tile size $\textit{T}^{out}_{\textit{K}}$=4. Therefore there will be 
Suppose there are total of two clusters that contain four PEs each. The four PEs in each cluster are spatially mapped onto dimension \textit{K} (column of \textit{A} and row of \textit{B}). Since two clusters are spatially mapped onto dimension \textit{N}, entire eight PEs are fully utilized as PEs 0--3 and 4--7 can compute $C_{0,0}$ and $C_{0,1}$ entries, respectively.
%in parallel via \textit{SpatialMap(1,1) \textit{K}}.
However, this non-tiled mapping does not provide the best performance. Hereafter, given any loop order, if the parallelism in the outer cluster is only on the innermost dimension (which is \textit{K} in \textless{}\textit{m}, \textit{n}, \textit{k}\textgreater{} loop order) and the tile sizes of \textit{both} outer dimensions (\textit{M} and \textit{N} here) are set to 1, we will call this \textbf{non-tiled mapping}.
~\autoref{fig:modeling_FLASH_GEMM_example} (b) shows a 2D tiling case when $\textit{T}^{out}_{\textit{M}}$=$\textit{T}^{in}_{\textit{M}}$=2, $\textit{T}^{out}_{\textit{N}}$=$\textit{T}^{in}_{\textit{N}}$=2, $\textit{T}^{out}_{\textit{K}}$=2, and $\textit{T}^{in}_{\textit{K}}$=1. Such mappings with non-unit tile sizes are called \textbf{tiled mapping}.
%In this case, there are 8 PEs$/\textit{T}^{out}_{\textit{K}}$=4 clusters that contain two PEs each. 
The tile size chosen $\textit{T}^{out}_{\textit{N}}$=2 yields under-utilization of clusters as
%As each cluster is spatially mapped along dimension \textit{N} with a size of $\textit{T}^{out}_{\textit{N}}$=2 in the outermost cluster via \textit{SpatialMap($\textit{T}^{out}_{\textit{N}}$,$\textit{T}^{out}_{\textit{N}}$) \textit{N}}, only \textit{N}$/\textit{T}^{out}_{\textit{N}}$=
only two out of four clusters can be mapped onto \textit{N} dimension, i.e., the last two clusters, Clusters 2 and 3, will be idle. This results in under-utilization of PEs 4--7 in the accelerator.
In the case shown in~\autoref{fig:modeling_FLASH_GEMM_example} (c), the tile size $\textit{T}^{out}_{\textit{N}}$ is determined by \textit{N}$/$(\textit{number of clusters}). 
The number of clusters in MAERI-style ({\sf TST\_TTS}) mapping is associated with the total number of PEs and the tile size of the dimension involving the last loop (which is $\textit{T}^{out}_{\textit{K}}$ in \textless{}\textit{m}, \textit{n}, \textit{k}\textgreater{} dataflow).
Hence, the optimal tile size $\textit{T}^{out}_{\textit{N}}$ can be calculated by $\textit{T}^{out}_{\textit{N}}$=\textit{N}$/$(\textit{P}$/\textit{T}^{out}_{\textit{K}}$) where \textit{P} denotes total number of PEs. \autoref{fig:modeling_FLASH_GEMM_example} (c) shows an example of 2D-tiled {\sf TST\_TTS-MNK} mapping when the appropriate tile sizes $\textit{T}^{out}_{\textit{M}}$=$\textit{T}^{in}_{\textit{M}}$=2, $\textit{T}^{out}_{\textit{N}}$=$\textit{T}^{in}_{\textit{N}}$=1, $\textit{T}^{out}_{\textit{K}}$=2, and $\textit{T}^{in}_{\textit{K}}$=1 are chosen for fully utilizing PEs and \emph{maximizing data reuse}.
We describe how we chose the tile sizes analytically in~\autoref{sec:flash}.
The key observation is that Clusters 0--3 compute partial outputs of four entries in matrix \textit{C} in parallel, while enabling data reuse of entries in matrix \textit{B}. While the idea is similar to tiling on temporal architectures, the mix of spatial and temporal data reuse has to be handled when tiling.
%The number of clusters in MAERI-style mapping is associated with the total number of PEs and the tile size of the dimension involving the last loop (which is $\textit{T}^{out}_{\textit{K}}$ in \textless{}\textit{m}, \textit{n}, \textit{k}\textgreater{} dataflow).
%Hence, the optimal tile size $\textit{T}^{out}_{\textit{N}}$ can be calculated by $\textit{T}^{out}_{\textit{N}}$=\textit{N}$/$(\textit{P}$/\textit{T}^{out}_{\textit{K}}$) where \textit{P} denotes total number of PEs. 

% \input{tables/flash_all_mappings}
%\subsection{Mapping \gemm on Spatial Accelerators}

\vspace{-2mm}
\subsection{\newmaestro cost model}
\label{sec:newmaestro}
Given a GEMM mapping described via the dataflow directives, we evaluate its performance on a target spatial accelerator using our cost model which we call \newmaestro.
\newmaestro builds upon a previously published open-source analytical cost model called \maestro~\cite{kwon2019understanding, maestro_web}
%which is an open-source analytical cost model targeting DNN operations on spatial accelerators.
\maestro receives a DNN model, accelerator hardware configuration, and mapping description %written in \maestro's data-centric directives 
as inputs, as shown in \autoref{fig:maestro_directives}.
It analytically (i.e., via detailed equations) analyzes the inputs and produces the expected runtime, 
number of buffer accesses, arithmetic intensity, NoC bandwidth requirement, and so on as outputs. It also reports the energy consumption based on energy of HW building blocks of accelerators (fixed point MAC units, SRAMs, buses, and so on) from CAD tools which are scaled based on the hardware configuration provided as input~\cite{kwon2019understanding}.
Its results have been validated against
the Eyeriss chip~\cite{chen2016eyeriss} and RTL simulations of MAERI~\cite{kwon2018maeri}.
%The output report provides more than 50 statistics that include expected runtime in cycles, number of buffer accesses, arithmetic intensity, NoC bandwidth requirement, and so on. 
%\maestro relies on an analytical method, i.e., detailed equations, to perform this analysis, making it orders of magnitude faster than cycle-accurate simulation~\cite{kwon2019understanding, maestro_web}. 
%For example, it takes 439ms to analyze the costs of running Resnet50 on an NVDLA-like accelerator, while providing 96\% accuracy in results compared to low-level hardware simulation results~\cite{kwon2019understanding}.
\maestro, however, does not accept BLAS operations as is, and requires users to map BLAS operations such as \gemm onto DNN operators (e.g., CONV2D). Our \newmaestro updates adds native BLAS kernel frontend in \maestro, which allows users to directly work on BLAS operations without converting them into DNN operators. \newmaestro leverages \maestro's backend for the performance and energy estimates.
\section{FLASH: A Mapping Explorer for \gemm}
\label{sec:flash}

\autoref{fig:FLASH_Overview_wide} shows an overview of our framework \flash for exploring \gemm mappings. We describe how we derive candidate tile sizes for the 
%five 
accelerators,  prune the search space using dataflow, hardware and tile sizes constraints, and evaluate the mappings using 
the \newmaestro cost model.

Given a \gemm dataflow for an accelerator-style, a \gemm workload and hardware configurations, exploring search space for all possible mappings is extremely compute-intensive if every feasible loop order, every cluster size, and every possible tile size are taken into consideration. Therefore, we propose a new search space reduction algorithm which systematically prunes the possible mappings by seeking the candidate (near optimal) tile sizes within the \flash mapping explorer.
% \input{algorithms/FLASH_generate_mapping_candidates}

%% TK: moved by Tushar from Section III
\betterparagraph{Dataflow and Hardware Constraints}
\flash honors the hardware constraints (expressed via the dataflow in \autoref{table:flash_mapping_comparison}) across the various accelerators, as described next. 
(i) The \gemm dimensions bound to the directives (i.e., loop order) is fixed except for
MAERI-style mapping where all loop orders are supported by the hardware. 
%(iv) Loop order between \textit{M} and \textit{N} dimensions of \gemm can always be swapped within an outer/inner cluster when \textit{K} dimension is used for accumulation.
%\squishend
(ii) Cluster size is fixed in every \gemm mapping except the MAERI-style mappping ({\sf TST\_TTS}). (iii) Tile size is a flexible input to any directive.

\betterparagraph{Candidate Mapping Selection}
Algorithm \ref{alg:flash_setup} in Appendix shows the pseudo-code for generating mapping candidates in \flash.
Given the type of accelerators (e.g., Eyeriss/NVDLA), hardware parameters (e.g., S1/S2 buffer sizes), and the sizes of \textit{M}, \textit{N}, and \textit{K} as inputs, \flash generates mapping candidates.
Based on the accelerator chosen by the user, \flash first determines three parameters -- dataflow directive order, all feasible loop orders, and all possible cluster sizes that can be explored in the mapping candidates based on the hardware constraints (line 1).
%These are based on the hardware constraints described in \autoref{subsec:modeling_spatial_accelerators} and  Table~\ref{table:flash_mapping_comparison}. After choosing these three parameters,
Next, \flash computes the candidate tile sizes using problem dimensions and hardware parameters (lines 7 and 8). These tile sizes are used to prune a huge number of possible mappings (lines 9 and 10).
%Finally, \flash generates the pruned mapping candidates descriptions as inputs to \newmaestro and analyzes the results of \newmaestro to choose the best \gemm mapping based on projected runtime.

\betterparagraph{Tile Size Selection}
Given an arbitrary accelerator and a large-scale \gemm workload, for each feasible loop order, all possible combinations of the tile sizes can be obtained by 
%performing six nested loops where each loop iterates over
using the maximum tile size acceptable for each dimension in outer and inner clusters. In most cases, each of the maximum tile sizes in the outer-cluster are equivalent to its actual dimension size \emph{if there are no other constraints}. We consider these as the baseline candidates. Three constraints are imposed to reduce the candidates further: (1) the inter-cluster tile sizes of the three matrices, \textit{A}, \textit{B} and \textit{C}, should fit within the S2 buffer;
%The tile sizes involved in the outer cluster and inner cluster have to be determined separately so those tile sizes can be used for a large-scale \gemm mapping. 
(2) the required memory for all clusters to be less than the S2 buffer; (3) the required memory for each PE to be less than the S1 buffer. In other words, the inner tile sizes must fit into the S1 buffer and must be a subset of the outer tile sizes (e.g., $\textit{T}^{\textit{in}}_{\textit{M}}<=\textit{T}^{\textit{out}}_{\textit{M}}$). Determining the candidate tile sizes is shown in lines 7--9 in Algorithm \ref{alg:flash_setup} in Appendix.

\emph{We show how these tile sizes can be calculated for one example}, the MAERI-style {\sf TST\_TTS} mapping with \textless{}\textit{m}, \textit{n}, \textit{k}\textgreater{} loop order (last column of Table \ref{table:flash_mapping_comparison}) using the constraints discussed above. %The result of this example is shown in the first column of Table \ref{table:tst_tts_mapping}. 
% Table \ref{table:all_mappings_tile_size} summarizes the candidate tile sizes for \emph{all} mapping schemes for all accelerators based on the same inequalities.
% \rev{Table \ref{table:all_mappings_tile_size} summarizes the candidate tile sizes for NVDLA- and MAERI-style mappings based on the same inequalities. The candidate tile sizes for other mapping schemes can be derived in a similar way. We do not list them because of space constraints but use them in our evaluations directly.}

For a MAERI-style mapping, note $\textit{T}^{out}_{\textit{N}}$ can be computed  as described in Section \ref{subsec:walk_through_example} because \textit{N} is spatially-mapped.  Hardware parameters such as S1/S2 buffer sizes must be considered to calculate near optimal candidate tile sizes $\textit{T}^{out}_{\textit{M}}$ and $\textit{T}^{out}_{\textit{K}}$ involved in the temporally-mapped dimensions \textit{M} and \textit{K}, respectively. Hence, given S2 and S1 buffer sizes, the tile sizes for outer cluster and inner cluster will be limited by the constraints in Equations \ref{eq:optimal_tile_size_L2} and \ref{eq:optimal_tile_size_L1}, respectively. We assume double-buffering for latency hiding in these calculations.
% \setlength{\belowdisplayskip}{0pt} \setlength{\belowdisplayshortskip}{0pt}
% \setlength{\abovedisplayskip}{0pt} \setlength{\abovedisplayshortskip}{0pt}
% \begin{equation}
% \scriptsize
% \label{eq:optimal_tile_size_L2}
% \textit{T}^{out}_{\textit{M}} \times \textit{T}^{out}_{\textit{K}} + \textit{T}^{out}_{\textit{K}} \times \bigg(\textit{T}^{out}_{\textit{N}} \times floor\Big(\frac{P}{\textit{T}^{out}_{\textit{K}}}\Big)\bigg) + \textit{T}^{out}_{\textit{M}} \times \bigg(\textit{T}^{out}_{\textit{N}} \times floor\Big(\frac{P}{\textit{T}^{out}_{\textit{K}}}\Big)\bigg) \leq \beta \times \frac{1}{2}
% \end{equation}
\begin{equation}
% \scriptsize
\footnotesize
\label{eq:optimal_tile_size_L2}
\textit{T}^{out}_{\textit{M}} \times \textit{T}^{out}_{\textit{K}} + \textit{T}^{out}_{\textit{K}} \times \bigg(\textit{T}^{out}_{\textit{N}} \times \Big(\frac{P}{\textit{T}^{out}_{\textit{K}}}\Big)\bigg) + \textit{T}^{out}_{\textit{M}} \times \bigg(\textit{T}^{out}_{\textit{N}} \times \Big(\frac{P}{\textit{T}^{out}_{\textit{K}}}\Big)\bigg) \leq \beta \times \frac{1}{2}
\end{equation}
\begin{equation}
% \scriptsize
\footnotesize
\label{eq:optimal_tile_size_L1}
\textit{T}^{in}_{\textit{M}} \times \textit{T}^{in}_{\textit{K}} + \textit{T}^{in}_{\textit{K}} \times \textit{T}^{in}_{\textit{N}} + \textit{T}^{in}_{\textit{M}} \times \textit{T}^{in}_{\textit{N}} \leq \alpha \times \frac{1}{2}
\end{equation}
where $\alpha$ and $\beta$ denote S1 buffer size and S2 buffer size, respectively.
For the candidate outer tile sizes for S2 buffer size, as the dimension \textit{N} is spatially mapped in the outer cluster-level, $\textit{T}^{out}_{\textit{N}}$ in Equation \ref{eq:optimal_tile_size_L2} can be replaced with \textit{N}$/$(\textit{P}$/\textit{T}^{out}_{\textit{K}}$).
Then if $\textit{T}^{out}_{\textit{M}}$ and $\textit{T}^{out}_{\textit{K}}$ are assumed to be equal, %solution to Equation \ref{eq:optimal_tile_size_L2_N_replace} is
the candidate tile sizes are shown in Equation \ref{eq:optimal_tile_size_L2_solution}.
We iteratively decrease the largest tile size when the tiles do not fit in the S2 buffer. Such corner cases might occur due to assumptions like $\textit{T}^{out}_{\textit{K}}$ and $\textit{T}^{out}_{\textit{M}}$ are the same.
\begin{equation}
% \scriptsize
\footnotesize
\label{eq:optimal_tile_size_L2_solution}
1 \leq \textit{T}^{out}_{\textit{M}} \leq \sqrt{\frac{\beta}{2}+N^{2}}-N\; , \; 1 \leq \textit{T}^{out}_{\textit{K}} \leq \sqrt{\frac{\beta}{2}+N^{2}}-N\; , \; \textit{T}^{out}_{\textit{N}}=\frac{\textit{N}\textit{T}^{out}_{\textit{K}}}{\textit{P}}
\end{equation}
If $\textit{T}^{in}_{\textit{M}}$ and $\textit{T}^{in}_{\textit{N}}$ are assumed to be equal, the solution to Equation \ref{eq:optimal_tile_size_L1} for S1 buffer size is shown in Equation \ref{eq:optimal_tile_size_L1_solution} (the tile size $\textit{T}^{in}_{\textit{K}}$ is always one in MAERI-style \textless{}\textit{m}, \textit{n}, \textit{k}\textgreater{} loop order).
\begin{equation}
% \scriptsize
\footnotesize
\label{eq:optimal_tile_size_L1_solution}
1 \leq \textit{T}^{in}_{\textit{M}} \leq \sqrt{\frac{\alpha+2}{2}}-1\; , \;  
1 \leq \textit{T}^{in}_{\textit{N}} \leq \sqrt{\frac{\alpha+2}{2}}-1\; , \; \textit{T}^{in}_{\textit{K}}=1
\end{equation}
% Table \ref{table:tst_tts_tile_size} summarizes the optimal tile sizes for six different loop order algorithms with {\sf TST\_TTS} mapping scheme based on the same inequalities.
For the tile sizes $\textit{T}^{out}_{\textit{M}}$, $\textit{T}^{out}_{\textit{K}}$, $\textit{T}^{in}_{\textit{M}}$ and $\textit{T}^{in}_{\textit{N}}$, the largest power of two (constrained by Equations \ref{eq:optimal_tile_size_L2_solution} and \ref{eq:optimal_tile_size_L1_solution}) result in better performance with respect to energy consumption and execution time.
Equations \ref{eq:optimal_tile_size_L2} and \ref{eq:optimal_tile_size_L1} can be used to find the candidate tile size for different mapping schemes such as {\sf STT\_TTS} and {\sf STT\_TST}.
% SR: I think is ok to cut.
%given any loop orders can also be computed by replacing the tile size of spatially-mapped dimension in the outer cluster with an appropriate variable. Also, the number of clusters (e.g, $P/T^{out}_{K}$ (=$P/\gamma$) in Equation \ref{eq:optimal_tile_size_L2}) is different for each mapping scheme. 
% \input{algorithms/FLASH_get_non_pruned_tile_sizes}
%\input{algorithms/FLASH_get_pruned_tile_sizes}
We avoid derivations for other mappings due to space constraints. The candidate tile sizes for other mapping schemes can be derived in a similar way. Table \ref{table:all_mappings_tile_size} in Appendix summarizes the candidate tile sizes for all mapping schemes for all accelerators based on the same inequalities.

\betterparagraph{Mapping Candidates Generation}
Once the candidate tile sizes are selected for each matrix dimension in the outer and inner clusters, all other tile sizes outside this range can be pruned. The tile sizes can then be combined with the chosen directive, loop order, cluster size selections to enumerate all the mapping candidates.
%, $\textit{T}^{out}_{\textit{M}}$, $\textit{T}^{out}_{\textit{N}}$, $\textit{T}^{out}_{\textit{K}}$, $\textit{T}^{in}_{\textit{M}}$, $\textit{T}^{in}_{\textit{N}}$, and $\textit{T}^{in}_{\textit{K}}$, as shown in Algorithm \ref{alg:flash_pruned_tile_sizes}. 
%This pruning scheme is highly beneficial to explore fewer search space. 
%As an example, when \textit{M}=256, \textit{N}=256 and \textit{K}=256 for {\sf TST\_TTS} mapping with \textless{}\textit{i}, \textit{j}, \textit{k}\textgreater{} loop order, the elapsed time for searching non-pruned 7,250,826,667 sets of tile size combination is 33415.27 (s), whereas the elapsed time for finding pruned 14,992,384 sets of tile size combination is only 27.75 (s). Hence, FLASH significantly decreases the number of mapping candidates by a factor of 483.63 and reduces the time for generating mapping candidates for \newmaestro by 99.9\% on a standard laptop.
%\insertFigure{FLASH_num_mappings_runtime}{Histogram of the projected runtime of the pruned mapping candidates for a NVDLA-style {\sf \textbf{STT\_TTS-NKM}} mapping For a \gemm of size (8192$\times$8192)$\times$(8192$\times$8192). In this case, \flash generates a total of 7387 mapping candidates. Each bin holds the uniform width (85.85 ms) of runtime and each bar represents the number of mapping candidates included in each bin.\vspace{-1mm}}

\betterparagraph{Mapping Candidates Selection using \newmaestro}
Given the pruned mapping candidates generated by Algorithm \ref{alg:flash_setup} in Appendix, \flash uses them as inputs to the \newmaestro cost model described earlier in \autoref{sec:newmaestro} and selects the best mapping based on the lowest projected runtime.

\section{Experimental Evaluation}
\label{sec:evaluation}
%Factors such as the sizes of matrices, spatially-mapped dimension, tile size, cluster size, and other hardware resources are critical for the performance of mappings. To address the impact of these factors precisely, we provide extensive comparisons of the mapping variants shown in Table \ref{table:flash_mapping_comparison}. % in Section \ref{sec:flash}.
In this section, we divide the evaluation into two parts.
First we demonstrate the impact of our framework by showing the effect of pruning in Section \ref{subsec:search_space_pruning} and tiling in Section \ref{subsec:impact_flash_tiling}. Second, we evaluate the accelerator architectures for different matrix shapes, loop orders and cluster sizes in Section \ref{subsec:evaluation_accelerators_workloads}.

\subsection{Evaluation Methodology}
\label{subsec:evaluation_methodology}
\betterparagraph{\gemm Workloads}
We select six distinct \gemm workloads, as listed in~\autoref{table:eval_workload}, inspired from several use cases to cover a wide variety of sizes and shapes~\cite{davis2016survey,bavier1970amesos2,kaagstrom1998gemm}.
The workload includes \gemm with large number of FLOPs (up to 549 GFLOPs) and different aspect ratios to show impact.
We also choose the shapes considering their applicability in real applications especially, tall-skinny, short-fat, square, rank-K update like problems. %for direct, iterative solvers and other \blas kernels.% in \gemm.
%The values in the \gemm workload do not impact the performance.
\begin{table}[ht]
\centering
\caption{The \gemm workloads we use for evaluations.}
\scalebox{0.9}{
\begin{tabular}{c|c|c|c|c|c|c}
\thickhline
\multirow{2}{*}{\begin{tabular}[c]{@{}c@{}}\textbf{Matrix}\\ \textbf{Dimension}\end{tabular}} & \multicolumn{6}{c}{\textbf{Workload ID}}                                   \\ \cline{2-7} 
                              & \textbf{I} & \textbf{II} & \textbf{III} & \textbf{IV} & \textbf{V} & \textbf{VI} \\ \hline
\textit{M}                    & 8192          & 1024          & 8          & 8          & 8192          & 512          \\ \hline
\textit{N}                    & 8192          & 1024          & 8          & 8192          & 8          & 256          \\ \hline
\textit{K}                    & 8192          & 8192          & 8192          & 1024          & 1024          & 256          \\ \hline \hline
GFLOPs                        & 549.8          & 8.59          & 0.001          & 0.067          & 0.067          & 0.03          \\ \thickhline
\end{tabular}
}
\label{table:eval_workload}
\end{table}

\betterparagraph{Mappings}
We use five mapping schemes following the style of the five accelerators, as shown in \autoref{table:flash_mapping_comparison}.
As MAERI-style mapping {\sf TST\_TTS} has the flexibility to use different loop order, we use \textless{}\textit{m}, \textit{n}, \textit{k}\textgreater{} loop order unless specified.
%The effect of different loop orders for all mappings are shown in~\autoref{fig:evaluation_loop_order} and discussed later.
We use the best performing candidate tile sizes computed using the method discussed in~\autoref{sec:flash} or their closest power of two that fit in S2 buffer.
We also select the best cluster size parameter ($\lambda$) for each mapping considering the micro-architectural constraints from accelerators that motivated each mapping (\autoref{table:flash_mapping_comparison}).

\betterparagraph{Hardware Configuration}
For a fair evaluation of the design of the architecture (as opposed to the instance of an architecture), we use the same hardware parameters (number of PEs, buffer sizes and NoC bandwidth) for each one of them targeting edge and cloud devices with a 2D PE array, as described in~\autoref{table:hardware_config}. We assume 1GHz clock and 28nm node. The configurations are based on those of previously proposed DNN accelerators~\cite{chen2016eyeriss, nvdla, jouppi2017datacenter} to show the performance of the accelerators under realistic settings.
\begin{table}[ht]
\centering
\caption{Hardware configurations we use for evaluations. We assume a 1 GHz clock for the accelerators at 28nm. Performance goal is based on the number of PEs and the clock rate.}
\scalebox{0.85}{
{
\begin{tabular}{c|c|c|c|c|c|c}
\thickhline
\textbf{ID} & \textbf{\begin{tabular}[c]{@{}c@{}}\# of \\ PEs\end{tabular}} & \textbf{\begin{tabular}[c]{@{}c@{}}S1\\ Size\end{tabular}} & \textbf{\begin{tabular}[c]{@{}c@{}}S2\\ Size\end{tabular}} & \textbf{\begin{tabular}[c]{@{}c@{}}NoC\\ BW\end{tabular}} & \textbf{\begin{tabular}[c]{@{}c@{}}Perf\\ FLOPS\end{tabular}} &
\textbf{\begin{tabular}[c]{@{}c@{}}Off-chip\\ Mem\end{tabular}} \\ \hline
Edge        & 256                                                           & 0.5 KB                                                            & 100 KB                                                            & 32 GB/s                                                          & 256 G  & DRAM                                                        \\ \hline
Cloud       & 2048                                                          & 0.5 KB                                                            & 800 KB                                                            & 256 GB/s                                                         & 2 T     
& HBM \\ \thickhline
\end{tabular}}}
\label{table:hardware_config}
\end{table}

% \begin{table}[ht]
% \centering
% \caption{Hardware configurations we use for evaluations. We assume a 1 GHz clock for the accelerators. Performance goal is based on the number of PEs and the clock rate.}
% \scalebox{0.76}{
% \begin{tabular}{c|c|c|c|c|c}
% \thickhline
% \textbf{ID} & \textbf{\# of PEs} & \textbf{S1 Buffer} & \textbf{S2 Buffer} & \textbf{NoC BW} & \textbf{Performance Goal} \\ \hline
% Edge & 256 & 0.5KB  & 100KB  & 32GB/s & 256 GFLOPS \\ \hline
% Cloud & 2048 & 0.5KB  & 800KB  & 256GB/s & 2 TFLOPS  \\ \thickhline
% \end{tabular}
% }
% \label{table:hardware_config}
% \end{table}
.

\betterparagraph{Hardware Modeling.}
The modeling of each accelerator is done by \newmaestro (\autoref{sec:newmaestro}), 
which uses \maestro's analytical equations and hardware backend~\cite{kwon2019understanding}. The S2 buffers are double-buffered to allow prefetches of the next tiles from memory while the current tiles are being computed. 
The reported energy in our evaluations is for the on-chip data accesses and movement, 
since the total off-chip data movement to fetch all operands and write all outputs remains similar across mappings~\cite{angshu2019timeloop}.

% \begin{comment}

% \betterparagraph{Validity Check of Mappings}
% We check the validity of each mapping description before we run based on the following rules:

% \squishlist
% \item R1: Only one cluster directive exists.
% \item R2: Only one spatial map directive exists at each cluster level.
% \item R3: The mapping size and offset parameters of each directive are identical.
% \squishend

% R1 and R2 are based on the 2D PE array of the accelerator we employ, which allows two-level parallelism in each dimension.
% R3 guarantees that the mapping performs all the operation exactly once without any replicated or omitted computation.

% \HK{I am not sure if we need the mapping validity check (or, sanity check) here since the evaluated mapping schemes never violate this. Maybe we need to discuss earlier and argue that the mapping schemes we select are valid styles?}

% \subsection{Sanity Checking/Testing Metric}
% All the mappings used for the evaluation have only one \textit{Cluster} directive since we assumed accelerators based on 2D PE arrays. There is only one \textit{SpatialMap} directive in each cluster level to exploit parallelism in one dimension at each cluster level. Also, for every \textit{TemporalMap} and \textit{SpatialMap}, their size and offset are always same to cover all multiplications only once. Therefore, the elements that should be multiplied are mapped into only one of the PEs exactly once with the mappings we used.
% \GJ{How to calculate number of MACs?}

% \end{comment}
% \input{tables/evaluation_buffer_access}

\subsection{Search Space Pruning \& Mapping Candidates Reduction}
\label{subsec:search_space_pruning}
We begin by evaluating search space pruning in \flash. 
%\flash pruning scheme is beneficial to explore a smaller search space.
Given \textit{M}=256, \textit{N}=256 and \textit{K}=256, a MAERI-style mapping with \textless{}\textit{m}, \textit{n}, \textit{k}\textgreater{} loop order, and hardware constraints such as S1 buffer size=0.5 KB, S2 buffer size=100 KB, NoC bandwidth=32 GB/s and total number of PE=256, the total number of possible tile size combinations would be 7,250,826,667 sets and would take $\sim$9.3 hours %33415.27 (s) 
to even generate the mapping candidates for \newmaestro if there are no further constraints.
In order to reduce the search space of tiling, we first came up with the tile size constraints for each dimension of the matrices which can be obtained by Equations \ref{eq:optimal_tile_size_L2_solution} and \ref{eq:optimal_tile_size_L1_solution}.
By effectively pruning the search space based on the candidate tile size constraints, we could obtain a total of 14,992,384 sets of tile size combinations which requires just 27.75 seconds to generate the candidate inputs.
% However, by effectively pruning the search space using the tile size selection
% % analytical tile sizes
% in \flash results in 14,992,384 sets of tile size combinations which requires just 27.75 seconds to generate the candidate inputs. 
In this instance, \flash decreases the number of mapping candidates by a factor of 483.63 and reduces the time for generating mapping candidates for \newmaestro by 99.9\% on a standard laptop. Similar results were observed for other workloads as well. We omit these experiments as evaluating the baseline candidates will require unreasonable amount of time. Note that this demonstrates the effect of pruning only based on the tile size. \flash also uses the hardware constraints to restrict the search space at the outer loop level which results in even better performance as the number of candidates grow even further.

\insertFigure{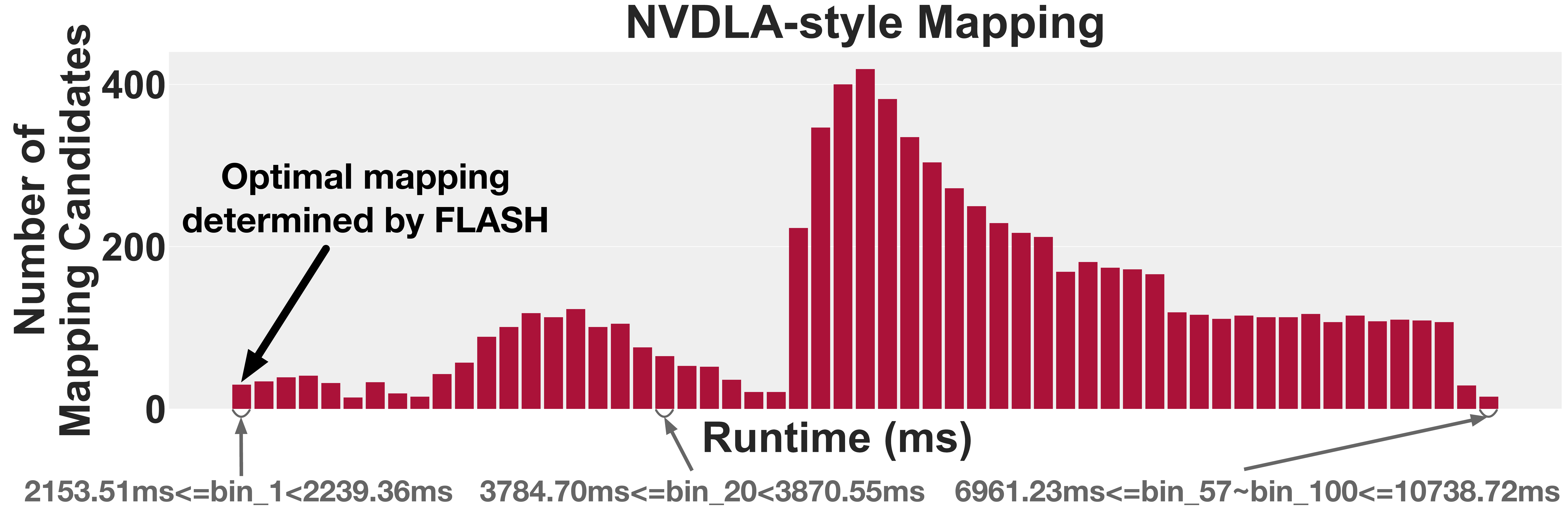}{Histogram of the projected runtime of the pruned mapping candidates for a NVDLA-style {\sf STT\_TTS-NKM} mapping for a \gemm of size (8192$\times$8192)$\times$(8192$\times$8192). In this case, \flash generates a total of 7,387 mapping candidates. Each bin holds the uniform width (85.85 ms) of runtime and each bar represents the number of mapping candidates included in each bin.}

\autoref{fig:FLASH_num_mappings_runtime} demonstrates effectiveness and importance of \flash with a different experiment. It shows an NVDLA-style \gemm mapping with a \textless{}\textit{n}, \textit{k}, \textit{m}\textgreater{} loop order for every cluster size, a total of 7,387 pruned mapping candidates are grouped into 100 bins based on their projected runtime values from \newmaestro. \flash selected mapping is in the bin with the lowest runtime. The figure also demonstrates that a ``bad'' mapping for this particular accelerator can be up to 4.02$\times$ slower than the best mapping. We also ran random sampling~\cite{angshu2019timeloop} and found that \flash consistently provided the same or better quality of mappings. \emph{To summarize, \flash reduces the time required for generating mapping candidates through the pruning phase while maintaining low projected runtime}. We plan to explore the multi-objective problem of choosing the mapping that is good in more than one quantity of interest in the future.

% the best projected runtime: We can't argue this since we don't have the comparison with the global optimum.
%8849.975/2196.435

\insertWideFigure{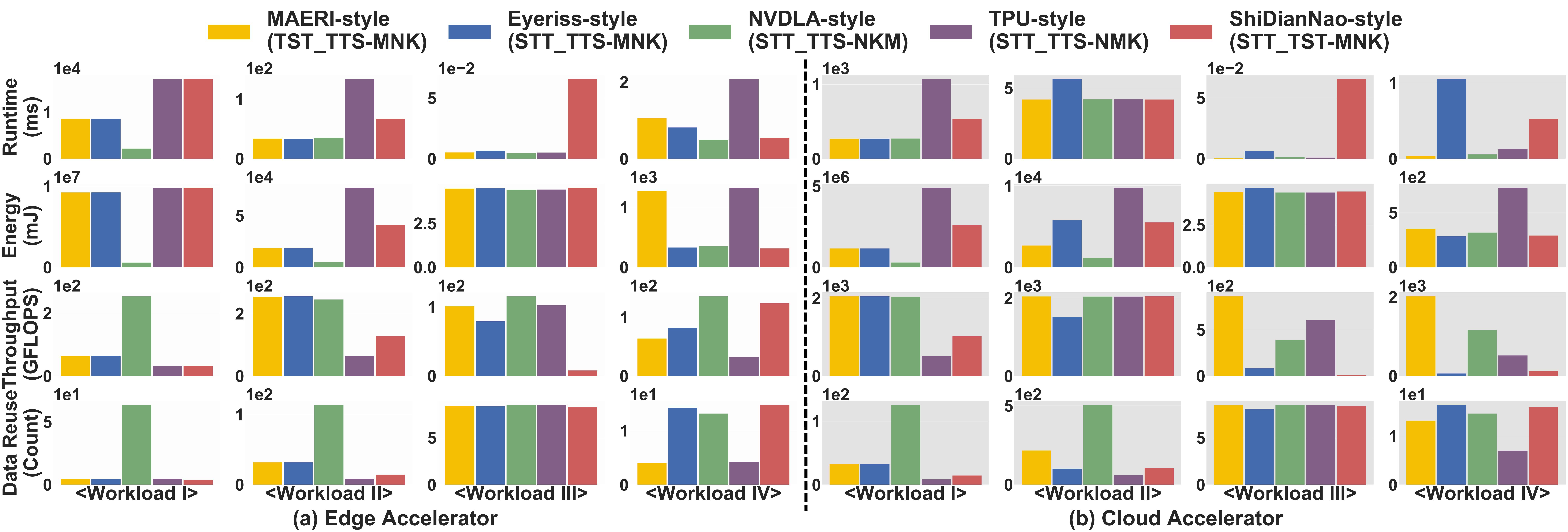}{\small
Runtime, energy, throughput, and data reuse of five mappings listed in~\autoref{table:flash_mapping_comparison} on (a) edge and (b) cloud accelerators. We apply the best performing cluster size ($\lambda$) and tile size (\textit{T}) for each mapping. The amount of data reuse is calculated by total number of S1 buffer accesses$/$total number of S2 buffer accesses. One can observe a correlation of data reuse to energy.}

\begin{table}[ht]
\centering
\caption{The impact of tiling on the number of buffer accesses, runtime, and energy. We use non-tiled ({\sf \textbf{NT}}) and tiled ({\sf \textbf{T}}) %{\sf \textbf{TST\_TTS}} 
MAERI-style mappings for different loop orders on workload VI and the edge configuration.}
\scalebox{0.615}{
\begin{tabular}{c|c|c|c|c|c|c|c|c|c}
\thickhline
\multirow{2}{*}{\textbf{\begin{tabular}[c]{@{}c@{}}\\Loop \\ Orders\end{tabular}}} & \multirow{2}{*}{\textbf{\begin{tabular}[c]{@{}c@{}}\\{\sf \textbf{NT}}/ \\ {\sf \textbf{T}}\end{tabular}}} & \multicolumn{3}{c|}{\textbf{Total S1 Access}}                                                                                                                                           & \multicolumn{3}{c|}{\textbf{Total S2 Access}}                                                                                                                                           & \multicolumn{2}{c}{\textbf{Performance}}                                                                                           \\ \cline{3-10} 
                                                                                  &                                  & \textbf{\begin{tabular}[c]{@{}c@{}}Matrix\\ A\end{tabular}} & \textbf{\begin{tabular}[c]{@{}c@{}}Matrix\\ B\end{tabular}} & \textbf{\begin{tabular}[c]{@{}c@{}}Matrix\\ C\end{tabular}} & \textbf{\begin{tabular}[c]{@{}c@{}}Matrix\\ A\end{tabular}} & \textbf{\begin{tabular}[c]{@{}c@{}}Matrix\\ B\end{tabular}} & \textbf{\begin{tabular}[c]{@{}c@{}}Matrix\\ C\end{tabular}} & \textbf{\begin{tabular}[c]{@{}c@{}}Run\\ time\\ (ms)\end{tabular}} & \textbf{\begin{tabular}[c]{@{}c@{}}Energy\\ (mJ)\end{tabular}} \\ \hline \hline
\multirow{2}{*}{\textbf{\textless{}\textit{m},\textit{n},\textit{k}\textgreater{}}}                                                      & {\sf \textbf{NT}}                      & 3.3E7                                                    & 6.6E7                                                    & 6.7E7                                                    & 2.6E5                                                      & 3.3E7                                                    & 2.6E5                                                      & 2.23                                                               & 570.02                                                         \\ \cline{2-10} 
                                                                                  & {\sf \textbf{T}}                       & 3.3E7                                                    & 3.3E7                                                    & 6.7E7                                                    & 2.6E5                                                      & 4.7E5                                                      & 6.5E5                                                      & \textbf{0.13}                                                               & \textbf{21.22}                                                          \\ \hline \hline
\multirow{2}{*}{\textbf{\textless{}\textit{n},\textit{m},\textit{k}\textgreater{}}}                                                      & {\sf \textbf{NT}}                      & 6.7E7                                                    & 3.3E7                                                    & 6.7E7                                                    & 3.3E7                                                    & 1.3E5                                                      & 2.6E5                                                      & 2.23                                                               & 570.02                                                         \\ \cline{2-10} 
                                                                                  & {\sf \textbf{T}}                       & 3.4E7                                                    & 3.3E7                                                    & 6.7E7                                                    & 1.1E6                                                     & 1.3E5                                                      & 1.1E6                                                     & \textbf{0.13}                                                               & \textbf{37.76}                                                          \\ \hline \hline
\multirow{2}{*}{\textbf{\textless{}\textit{m},\textit{k},\textit{n}\textgreater{}}}                                                      & {\sf \textbf{NT}}                      & 3.3E7                                                    & 6.7E7                                                    & 6.7E7                                                    & 2.6E5                                                      & 3.3E7                                                    & 3.3E7                                                    & 1.31                                                               & 1132.31                                                        \\ \cline{2-10} 
                                                                                  & {\sf \textbf{T}}                       & 3.3E7                                                    & 3.3E7                                                    & 6.7E7                                                    & 2.6E5                                                      & 4.7E5                                                      & 2.6E5                                                      & \textbf{0.13}                                                               & \textbf{14.61}                                                          \\ \hline \hline
\multirow{2}{*}{\textbf{\textless{}\textit{n},\textit{k},\textit{m}\textgreater{}}}                                                      & {\sf \textbf{NT}}                      & 6.7E7                                                    & 3.3E7                                                    & 6.7E7                                                    & 3.3E7                                                    & 1.3E5                                                      & 3.3E7                                                    & 1.31                                                               & 1132.31                                                        \\ \cline{2-10} 
                                                                                  & {\sf \textbf{T}}                       & 3.4E7                                                    & 3.3E7                                                    & 6.7E7                                                    & 6.0E5                                                      & 1.3E5                                                      & 2.6E5                                                      & \textbf{0.13}                                                               & \textbf{14.61}                                                          \\ \hline \hline
\multirow{2}{*}{\textbf{\textless{}\textit{k},\textit{m},\textit{n}\textgreater{}}}                                                      & {\sf \textbf{NT}}                      & 3.3E7                                                    & 3.3E7                                                    & 6.7E7                                                    & 2.6E5                                                      & 1.3E5                                                      & 3.3E7                                                    & 1.31                                                               & 571.12                                                         \\ \cline{2-10} 
                                                                                  & {\sf \textbf{T}}                       & 3.3E7                                                    & 3.3E7                                                    & 6.7E7                                                    & 2.6E5                                                      & 1.3E5                                                      & 1.1E6                                                     & \textbf{0.13}                                                               & \textbf{24.26}                                                          \\ \hline \hline
\multirow{2}{*}{\textbf{\textless{}\textit{k},\textit{n},\textit{m}\textgreater{}}}                                                      & {\sf \textbf{NT}}                      & 5.0E7                                                    & 3.3E7                                                    & 6.7E7                                                    & 1.6E7                                                    & 1.3E5                                                      & 3.3E7                                                    & 1.31                                                               & 852.26                                                         \\ \cline{2-10} 
                                                                                  & {\sf \textbf{T}}                       & 3.3E7                                                    & 3.3E7                                                    & 6.7E7                                                    & 2.6E5                                                      & 1.3E5                                                      & 6.5E5                                                      & \textbf{0.13}                                                               & \textbf{15.44}                                                          \\ \thickhline
\end{tabular}
}
\label{table:non_tiled_vs_tiled}
\end{table}

\subsection{Impact of \flash Tiling}
\label{subsec:impact_flash_tiling}
We demonstrate the impact of tiling chosen by \flash for MAERI-style mapping for different loop orders. 
Table \ref{table:non_tiled_vs_tiled} compares the number of buffer accesses of the two scratchpads for non-tiled and tiled MAERI-style mappings running workload VI with different loop orders. The number of total S2 accesses with tiled mappings was significantly smaller compared to that of non-tiled mappings for every loop order because the tiled version exploits data reuse using different strategies based on the loop order. \textbf{\textit{Overall tiling reduces runtime by 94\% and energy reduces by 96\% for the \textless{}\textit{m}, \textit{n}, \textit{k}\textgreater{} loop order}}. Similar improvements are seen with other loop orders. Since S2 scratchpad accesses consume significantly more energy compared to S1 access or computation, the reduction in S2 accesses leads to dramatic improvements in energy costs.
%Since we apply the same tile sizes and  % while the non-tiled versions do not reuse data in S1 scratchpad in PEs.
The runtime is also improved by reducing the amount of traffic between a PE and the S2 scratchpad, reducing the on-chip communication delay of an accelerator.
The impact of tiling is more significant than that of loop order as 91.25\% runtime reduction is observed by tiling on average while loop orders with the best and the worst runtime show 0.8\% difference in runtime within tiled mappings. Similar results were observed for tiling for other accelerators as well. Tiled versions with the tile size chosen by \flash were used for all mappings in the rest of the paper.

\subsection{Evaluation of Accelerators and Workloads}
\label{subsec:evaluation_accelerators_workloads}
The reduced search space based on the tile sizes and hardware constraints with \flash in addition to having a fast analytical model in \newmaestro allows us to study a larger space with five accelerators, two configurations, and six workloads for the pruned loop order, cluster size and tile sizes. We summarize these results in this sub-section.

\betterparagraph{The Impact of Matrix Shape}
~\autoref{fig:evaluation_matrix_shape} shows the runtime, energy, throughput, and data reuse of five mappings for four selected workloads 
%listed in~\autoref{table:eval_workload} 
to show the impact of matrix sizes and shapes.
%on the performance of each mapping scheme.
\textit{These experiments use a fixed loop order for fair comparison.}
NVDLA-style mapping does better than other mappings, 81.0\% lower runtime and 93.8\% less energy on the edge accelerator for square matrices (workload I).
However, on the cloud accelerator, all the mappings except ShiDianNao-style mapping achieve near-peak throughput (2 TFLOPS) and similar runtime. NVDLA-style mapping results in 87.8\% less energy, on average across mappings. \emph{When loop orders are fixed, NVDLA-style mapping scheme is the best among five mapping schemes for the square matrices (workload I)}.

Such results on square matrices are mainly based on two factors; loop order and tile size.
The loop order with \textit{K} at the inner-most position requires data tiles on both matrices \textit{A} and \textit{B} because both of them are coupled with dimension \textit{K}.
That is, it is hard to leverage any \textit{data reuse} if we place dimension \textit{K} at the inner-most position, which leads to bad energy efficiency.
The second aspect is the tile size that determines the balance between computation and communication delay in combination with loop order and NoC bandwidth~\cite{guirado2019understanding}.
If the communication delay for data tile fetch is longer than the computation delay of computation tiles on sub-clusters (or, PEs), the communication latency hiding fails, which leads to significant runtime increase.
We observe such cases on edge accelerators running MAERI-, Eyeriss-, and TPU-style mappings; they show near-peak bandwidth when the NoC bandwidth is larger in the cloud accelerator.

Short-and-fat matrices in workloads II and III (i.e., \textit{K} $>>$ \textit{M} and \textit{N}) show a different trend based on the skewness of the aspect ratio of matrices.
For the workload II, with aspect ratio of 1:8 between \textit{M}/\textit{N} and \textit{K}, MAERI-style and Eyeriss-style mappings provide the lowest runtime, which is 57.1\% less compared to other mappings, on average.
However, based on the impact of loop order, the mappings requires 73.6\% more energy compared to NVDLA-style which is the most energy efficient mapping across edge and cloud accelerators.
The energy consumption of Eyeriss-style mapping on cloud accelerator is larger than that on edge accelerator because the tiling parameters are changed and it affects the degree of spatial reuse via multi-casting.

Workload IV is \gemm of a short-and-fat matrix \textit{A} and a tall-and-skinny matrix \textit{B}. We observe NVDLA-style and MAERI-style mappings provide the lowest runtime on edge and cloud accelerators, respectively.
\emph{This use case demonstrates the success of workload and architecture-aware tiling strategy}. The large skew in dimension (\textit{M}:\textit{N}=1:1024) results in an extreme tiling strategy for MAERI-style mapping that maximizes data reuse on the smaller matrix \textit{A}, which  significantly reduces the number of expensive S2 buffer accesses.

Next, we identify that the amount of data reuse directly impacts energy usage with a negative correlation.
This is because the number of S2 accesses dominate the on-chip energy~\cite{kwon2019understanding}, and data reuse indicates the reduction of the number of S2 accesses.
We also observe that high data reuse counts led to high throughput across workloads except workload III on ShiDianNao-style mapping. This is because data reuse and throughput do not have a direct correlation. Also, an output stationary accelerator is not an ideal choice when the size of output matrix $C$ is small as workload III.

We observe that data reuse dramatically differs between mappings depending on the matrix sizes and shapes. For example, NVDLA-style mapping was better suitable for workload I than III (7.5$\times$ more data reuse).
However, the trend is not the same for all the mapping styles. For MAERI-style mapping, we observe the opposite trend from NVDLA-style mapping; 1.9$\times$ more data reuse on workload III than workload I. This is partially due to microarchitecture differences, but fixed loop orders also play a role. Such results show the relationship between mapping and workload is not trivial, which requires consideration of workload, mapping, and architecture. Our \flash will allow architects to systemically evaluate all the options.

\emph{NVDLA-style performs the best across the workloads and accelerators}, providing 69.8\% lower runtime and 92.3\% less energy, on average. However, the non-square workloads prefer different mappings, which implies that no single mapping is ideal for all the workloads.
\flash enables adapting the mappings for such workloads and selects the best performing mapping for each workload. This can provide 4.4\% additional runtime and 0.7\% energy improvements compared to the mapping (NVDLA-style) optimized for the average case. In the cloud configuration, \emph{MAERI-style mapping achieves the peak FLOPS in three out of four workloads shown here.} The results could change significantly if we fully exploit the flexibility in loop order.

\insertFigure{evaluation_loop_order}{Performance comparison of the mapping variants by varying all feasible loop orders for MAERI-style ({\sf TST\_TTS}) mapping on two workloads (IV and V) and two accelerators (edge and cloud).}
\betterparagraph{The Impact of Loop Order}
~\autoref{fig:evaluation_loop_order} presents the runtime and energy cost of all the evaluated mappings varying six different loop orders for MAERI-style mapping across workloads IV and V.
\emph{We observe the loop order has significant impact on both runtime and energy in different degrees depending on the mapping scheme}.
% For example, {\sf STT\_TTS} mapping with the clustering parameter ($\lambda$) of one reduces 99.6\% of runtime when it switched from \textless{}\textit{n}, \textit{m}, \textit{k}\textgreater{} loop order to \textless{}\textit{m}, \textit{n}, \textit{k}\textgreater{} at the cost of 23.7\% extra energy on workload V on the cloud accelerator.
For example, for the workload IV, energy usage reduces by 76.5\% on MAERI-style mapping (edge configuration) when switched from \textless{}\textit{m}, \textit{n}, \textit{k}\textgreater{} to \textless{}\textit{n}, \textit{m}, \textit{k}\textgreater{} loop order.
Furthermore, on the cloud configuration, MAERI-style mapping with \textless{}\textit{m}, \textit{k}, \textit{n}\textgreater{} and \textless{}\textit{n}, \textit{k}, \textit{m}\textgreater{} loop orders on workload IV results in 0.03 ms and 0.52 ms of runtime, respectively.
The trend reverses in workload V because workloads IV and V are transposes.

% We observe exactly opposite trends in the other workload IV because workloads IV and V are transposes, and we changed loop order between dimensions \textit{M} and \textit{N}, which is consistent with interchangeability rules we discussed in Section \ref{subsec:modeling_spatial_accelerators}.
% However, the results on workload V show the opposite preference.
%, which is consistent with interchangeability rules we discussed in Section \ref{subsec:modeling_spatial_accelerators}.

The preference to loop orders differs by the mapping schemes and workloads. For example, MAERI-style mapping with \textless{}\textit{m}, \textit{n}, \textit{k}\textgreater{} loop order on workload IV and cloud accelerator achieves 0.03 ms of runtime while Eyeriss-style with \textless{}\textit{m}, \textit{n}, \textit{k}\textgreater{} achieves 1.05 ms of runtime. %The preference can be reversed when the workload changes; the Eyeriss-style mapping has opposite preferred loop order when we change workload from IV to V.
The preference toward loop order based on workload and mapping scheme implies that the flexibility in loop order can provide significant runtime and energy benefits.
Across all the mappings and workloads IV and V on the edge accelerator, we observe that \emph{flexible loop order selected from \flash provides 49.9\% runtime and 49.7\% energy reduction} across mapping schemes compared to the average-workload-optimized fixed loop order of each mapping scheme.
\emph{On the cloud accelerator, the potential benefits were more significant on runtime (85.6\% improvements) than energy benefits (26.2\%) on average}.
We also swept the cluster size across the accelerators. We do not show results, in the interest of space, but found that it affects utilization, which in turn affects runtime and energy (up to 42\% in our results).

% \betterparagraph{The Impact of Cluster Size}
% From the results in~\autoref{fig:evaluation_loop_order}, we also observe the cluster size as another major factor that affects the overall performance of a mapping.
% For example, the {\sf STT\_TTS} mappings with different clustering parameters ($\lambda$) with \textless{}\textit{m}, \textit{n}, \textit{k}\textgreater{} on workload V and the edge accelerator consumed different amount of energy without monotonic trends with the cluster size.
% The most energy-consuming cluster size required 42.5\% more energy compared to the most energy-efficient cluster size.
% As we can observe, the relationship between cluster size, performance and energy is not trivial, which motivates a systematic approach like \flash to optimize cluster size with other factors.

% \insertFigure{evaluation_GEMM_MLP_MNIST_Accelerators_cols}{\rev{Performance comparison of the mapping variants on four \gemm workloads (i.e., fully-connected layers) involving Deep Neural Networks and edge accelerator. FC layers 1, 2, 3 and 4 performs \gemm of size (128$\times$784)$\times$(784$\times$512), (128$\times$512)$\times$(512$\times$256), (128$\times$256)$\times$(256$\times$128) and (128$\times$128)$\times$(128$\times$10), respectively.} \GM{Added Figure 10}}

\insertFigure{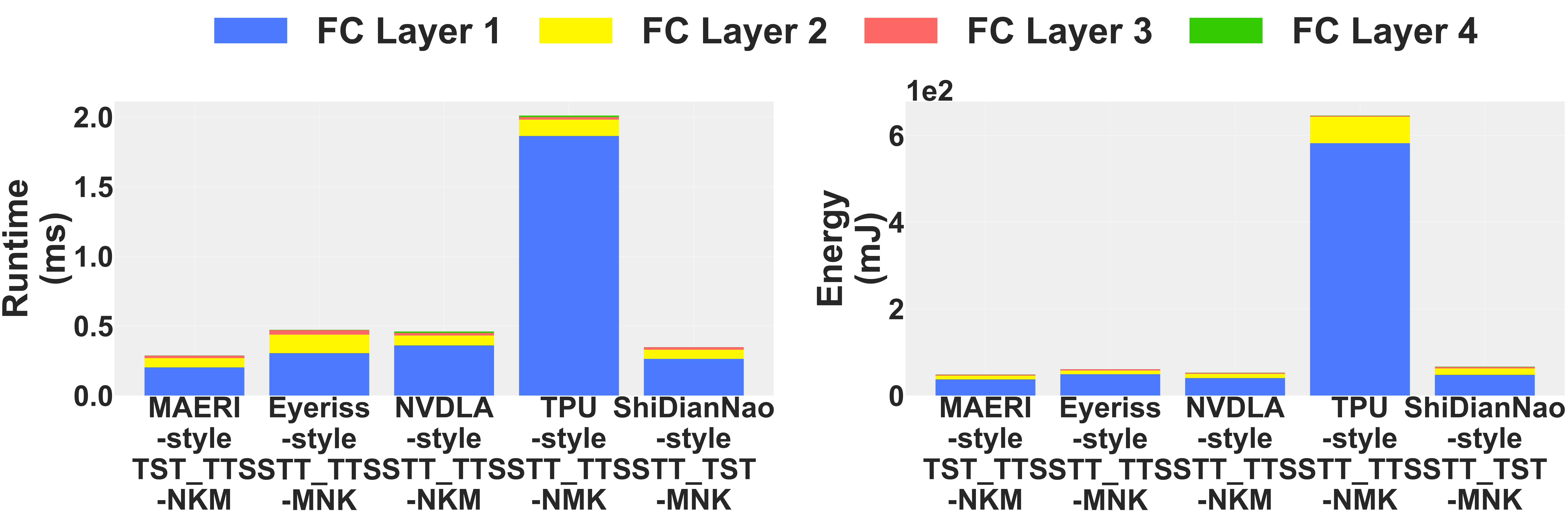}{Performance comparison of the mapping variants on four \gemm workloads (i.e., fully-connected layers) involving Deep Neural Networks and edge accelerator. A fully-connected layer performs \gemm of size (batch size$\times$\# of nodes in current layer)$\times$(\# of nodes in current layer$\times$\# of nodes in the next layer).}

% \insertFigure{evaluation_GEMM_MLP_MNIST_Accelerators_rows}{Performance comparison of the mapping variants on Deep Neural Networks.}

\betterparagraph{Performance on Deep Neural Networks}
As \gemm accounts for approximately 90\% of the total number of operations while training/testing DNNs such as fully connected Multi-Layer Perceptron (MLP), we also evaluated \gemm operations for inference of MLP model on different accelerators. The MLP model we used consists of an input layer, three hidden layers and an output layer, and 512, 256 and 128 nodes for each hidden layer. ~\autoref{fig:evaluation_GEMM_MLP_MNIST_Accelerators_cols} shows the runtime and energy cost across the various mappings for performing \gemm operations involved in inference of DNNs.

In general, batch data processing is used for DNN training/inference. Therefore \gemm kernel is used on batched data between one layer to the next. For example, using a real-world MNIST image dataset, a fully-connected (FC) layer 1 shown in \autoref{fig:evaluation_GEMM_MLP_MNIST_Accelerators_cols} connects 28$\times$28=784 nodes in input layer to 512 nodes in the first hidden layer. Hence, FC layer 1 corresponds to the first GEMM operation that multiplies an input matrix of size (128$\times$784) and a weight matrix of size (784$\times$512) where the batch size is set to 128. Similarly, FC layer 4 computes GEMM of size (128$\times$128)$\times$(128$\times$10) to connect 128 nodes in the third hidden layer to 10 nodes (classes) in output layer.
With these four GEMM workloads we used for four FC layers in MLP model. In these use cases, MAERI-style and ShiDianNao-style mappings provide lower runtime and MAERI-style and NVDLA-style provide higher energy efficiency relative to the other mappings. %Note that if matrices \textit{A} and \textit{B} are swapped, the performance results will be different as the spatially mapped (i.e., parallelized) dimension in each mapping will be changed.

% In our experiments, we measured the performance of conducting inference for MLP with a real-world image dataset. 
% The experimental results demonstrate that the size and shape of matrices are the critical factors that affects performance of GEMM mappings.
% through matrix transposition
\betterparagraph{Summary}
We summarize our key observations based on all the experiments:
%Based on the observations from our evaluation, we conclude the followings:
\squishlist
{\item \flash reduces the number of mapping candidates that needs to be considered by the analytical model by up to 480$\times$ and still finds a mapping that reduces runtime.}
{\item The new tiling strategy for spatial accelerators significantly reduces runtime and energy usage.}
{\item The evaluation of accelerators on several workloads show loop order is critical for runtime and energy in different degrees for each mapping scheme and workload. Matrix shapes also play an important role.}
%{\item No single loop order is ideal for all the mapping schemes and workloads.}
%{\item A systematic approach using mapping scheme, loop order, and cluster sizes with \flash can help accelerator designers and programmers to significantly improve the performance and efficiency of an accelerator.}
{\item Flexible mapping accelerators (e.g., MAERI~\cite{kwon2018maeri}), with a framework like \flash can provide significant runtime and energy improvements compared to the best mapping optimized for the average case, showing 65.3\% runtime and 30.1\% energy improvements on average.}
%from our evaluation results}.
\squishend
\section{Related Work}
\label{sec:related_works}
% \colonparagraph{Spatial DNN Accelerators}
% Spatial accelerators, an instance of the dataflow architecture~\cite{dennis1974preliminary,culler1986dataflow} %suggested in 70-80s,
% has become one of the most popular architectures for DNN accelerators since DNN applications have computation graphs without control flows and ample data-reuse opportunities, which are preferred by dataflow architectures, and the dataflow architecture requires lighter-weight hardware compared to general purpose architectures.
% ShiDianNao~\cite{du2015shidiannao} is a DNN accelerator optimized for exploiting data reuse across sliding windows of convolution.
% Eyeriss~\cite{chen2016eyeriss} first highlighted the impact of dataflows in DNN accelerators and proposed an energy-efficient row-stationary dataflow style.
% Flexflow~\cite{lu2017flexflow} is one of the early flexible accelerators that support three different dataflow styles to adapt to various DNN layers.
% MAERI~\cite{kwon2018maeri} is a fully flexible-dataflow DNN accelerator that support any dataflow style via flexible interconnect.
% Tangram~\cite{gao2019tangram} and Simba~\cite{shao2019simba} explored the idea of deploying multiple 2D PE arrays within a chip and a package, respectively. These work do not focus on \blas operations.
%FPGA-based accelerators also ~\cite{sharma2016high, kala2019high, li2020heterohalide}
% \TK{I removed the para on Spatial accelerators as it overlaps with background. Instead I think we need some text on the following}

\colonparagraph{Mappers for Spatial Accelerators}
% \TK{@Prasanth could you add some text on
% related work on mappers in AutoTVM, dmazeruner, Timeloop etc. Borrow some text from marvel paper. keep it short}
Most prior works focus on 
%Prior work~\cite{zhao2019mRNA} focused on
developing mappers specific to their architectures. For example, mRNA~\cite{zhao2019mRNA} for MAERI~\cite{kwon2018maeri}, Auto-TVM~\cite{Chen:2018:TAE:3291168.3291211} for the \gemm core of the VTA architecture~\cite{DBLP:journals/micro/MoreauCVRYZFJCG19} limiting their applicability to generic spatial accelerators.
Mapping optimizers such as Interstellar~\cite{interstellar}, dMazeRunner~\cite{DMazeRunner} are specific to convolutions and
fix certain aspects of the dataflow (such as choice of parallel loops and loop order), constraining the search space.
%are specific to the dataflow for convolutions \SR{and do not deal with the tiling and pruning aspects}.
To the best of our knowledge, Timeloop~\cite{angshu2019timeloop} is the only framework that considers all aspects of a mapping for \gemm kernels on a flexible spatial accelerator. 
However, it employs either an exhaustive linear search or a random sampling-based heuristic.
%to explore the search space.

% \colonparagraph{Mappers for FPGA}
% Past work on FPGA hardware vary from \blas/LAPACK implementations~\cite{merchant2016accelerating,merchant2017accelerating,de2019fblas}, dense linear algebra kernels~\cite{underwood2004closing,kestur2010blas,zhuo2008high,zhuo2005design,zhuo2005high}, Cannon's \gemm \cite{gorlani2019opencl}, QR decomposition~\cite{wang2009truly}, least squares solvers~\cite{yang2009fpga}. 

% There are application level studies such as finite difference solvers~\cite{sano2007systolic,giefers2014accelerating}, iterative and direct solvers~\cite{morris2005fpga, zhang2012portable}, Lattice Boltzman method \cite{sano2007fpga} and works that include multiple such kernels~\cite{zohouri2016evaluating, meyer2020evaluating}. Entire book has been devoted the topic of reconfigurable computing~\cite{gokhale2006reconfigurable}. In addition, programming considerations have been explored by several authors~\cite{ben2019stateful,watanabe2020design,aliaga2017sycl}. 
% While informed by this broad field, the focus on linear algebra kernels mapped over multiple spatial ASIC accelerators is novel to this work.
% with the exception of two papers which we highlight next.
% in the next paragraph. Moreover, though \newmaestro models ASICs, the \flash mappings and corresponding insights for various \gemm dimensions can also apply to spatial accelerators overlaid 
% on FPGAs and is ripe for future work.
% The primary difference from the FPGA work is the ability to use ASICs built for DNNs for CSE applications as well.

\colonparagraph{\gemm Accelerators}
Pedram et al.,~\cite{pedram2011high} proposed a \gemm accelerator called Linear Algebra Core (LAC) which consists of a $\sqrt{P}\times \sqrt{P}$ 2D grid of PEs. 
% Assume that matrices \textit{A} and \textit{B} are partitioned into (\textit{M}$/\textit{T}_{\textit{M}}$)$\times$(\textit{K}$/\textit{T}_{\textit{K}}$) tiles and (\textit{K}$/\textit{T}_{\textit{K}}$)$\times$(\textit{N}$/\textit{T}_{\textit{N}}$) tiles, respectively. Here, $\textit{T}_{\textit{N}}$ must be equal to $\sqrt{P}$, and $\textit{T}_{\textit{M}}$ is further divided by $\sqrt{P}$ so that each of $\sqrt{P}\times \sqrt{P}$ PEs can read two elements, one from each tile of ($\sqrt{P}\times \textit{T}_{\textit{K}}$) in \textit{A} and ($\textit{T}_{\textit{K}}\times \sqrt{P}$) in \textit{B} to perform MAC operation in a systolic array.
LAC accelerator is based on SUMMA algorithm~\cite{van1997summa}. 
which can be considered as a tiled \gemm with \textless{}\textit{k}, \textit{m}, \textit{n}\textgreater{} loop order with specific multicast between rows and columns of PEs.
Pedram et al.,~\cite{pedram2012codesign} further proposed a multi-core like \gemm accelerator called Linear Algebra Processor (LAP) with multiple LACs and tiled matrices \textit{A} and \textit{B}. 
Each LAC acquires a disjoint tile $\textit{A}_{\textit{LAC\_id},\textit{k}}$ and a shared tile $\textit{B}_{\textit{k},\textit{j}}$ so as to ensure that all the LACs in LAP are able to perform \gemm simultaneously.
However, LAP is limited to the SUMMA algorithm and its mapping style.
In contrast, we focus on several accelerators and map several \gemm variations to them.

\colonparagraph{Auto-tuning/Tiling for \blas}
% Auto-tuning approaches such as ATLAS~\cite{whaley2001automated}, SPIRAL~\cite{4536398} are popular in generating optimized codes for \blas.
%over variety of platforms. 
% Ideas used by Atlas library is still very popular for autotuning \blas libraries \cite{whaley2001automated}. 
%For instance, SPIRAL focuses on auto-tuning of compiler-generated codes. ATLAS uses auto-tuning techniques for 
%%loop unrolling, register blocking, cache blocking (a.k.a 
%loop tiling optimizations, and also for exploring recursion-based alternative implementation choices. 
%The broader ideas of ATLAS to choose between block based algorithms for portability and specific parameter tuning are also described as self-adapting linear algebra \cite{demmel2005self}.
Auto-tuning approaches such as ATLAS~\cite{whaley2001automated}, SPIRAL~\cite{4536398} are popular in generating efficient, optimized, and portable codes for \blas operations over a variety of platforms. 
% Ideas used by ATLAS library is still very popular for auto-tuning \blas libraries \cite{whaley2001automated}. 
For instance, SPIRAL focuses on auto-tuning of compiler-generated codes. ATLAS uses auto-tuning techniques for 
%%loop unrolling, register blocking, cache blocking (a.k.a 
loop tiling optimizations, and also for exploring recursion-based alternative implementation choices. 
The broader ideas of ATLAS to choose between block based algorithms for portability and specific parameter tuning are also described as self-adapting linear algebra \cite{demmel2005self}. 
Instead of generating different micro-kernels to choose tiling as they do, we use a cost model to evaluate the variants. 
Recent work has explored tiling and batching of \gemm on
GPUs~\cite{li2019coordinated}. Libraries like CUTLASS~\cite{kerr2017cutlass} use a hierarchical tiling for a temporal architecture.
%
%Specific to \gemm, Gunnels et al.~\cite{gunnels2001flame} and Goto et al.~\cite{goto2008anatomy} are two approaches based on building block kernels for \gemm. Block based \cite{dayde1994parallel} or recursive approaches~\cite{gustavson1998superscalar} for \gemm have been popular for decades now. 
The tiling approach we use is similar to the block based methods~\cite{whaley2001automated} and a variation of Goto et al.~\cite{goto2008anatomy} as the accelerator hardware is far simpler (e.g., no TLB). An analytical derivation of tile size is possible in this work as we were able to focus on few hardware parameters that is common among all the accelerators. 
As accelerators diverge we might have adapt some of the techniques from the auto-tuning approaches as well.
Most of these past work limit themselves one popular loop order for CPU architectures. 
We derive tile sizes for all loop orders instead of one popular loop order, as the architectures are evolving.
\section{Conclusion}
\label{sec:conclusion}
We develop a framework \flash for evaluating spatial accelerators via the \gemm kernel.
%We demonstrate it with the \gemm kernel. 
\flash considers all aspects of the mapping (tile sizes and dataflow for temporal, spatial and spatio-temporal reuse) within the microarchitectural constraints from the accelerator (number of PEs, NoC capability, buffer hierarchy), prunes the search space based on these constraints and additional heuristics, to 
%automatically 
produce optimized mappings for \gemm. The experimental results show the importance of different co-design considerations computer architects and algorithm designers have to take.
This comprehensive study lays the first steps for a heterogeneous HPC node with these accelerators that can address the need of machine learning applications and computational science applications.
The ideas in this work could lead to future work studying other dense and sparse ML and CSE kernels over accelerators.

\ifCLASSOPTIONcaptionsoff
  \newpage
\fi


\bibliography{references.bib}{}
\bibliographystyle{IEEEtran}

% \input{sections/bios.tex}

% \newpage
\appendix
We include finer level details in this appendix. 

\begin{algorithm*}[ht]
% \scriptsize
\footnotesize
\SetInd{0.45em}{0.45em}
\SetAlgoLined
\DontPrintSemicolon
\KwIn{\textit{M, N, K}: matrix dimensions, \textit{Arch}: type of accelerator, \textit{P}: total number of PEs, $\alpha$: size of S1 buffer, $\beta$: size of S2 buffer, $\lambda$: size of cluster}
\KwOut{Mapping\_Candidates: mapping descriptions of all mapping candidates}

dataflow\_candidates $\leftarrow$ \textbf{get\_candidate\_LoopOrders\_ClusterSzs}(\textit{Arch}, \textit{P});\\
\textit{num\_LoopOrders} $\leftarrow$ \textbf{get\_num\_LoopOrders}(dataflow\_candidates);\\ \textit{num\_ClusterSzs} $\leftarrow$ \textbf{get\_num\_ClusterSzs}(dataflow\_candidates);\\

% \textit{num\_LoopOrders} $\leftarrow$ dataflow\_candidates.shape[0]; \textit{num\_ClusterSzs} $\leftarrow$ dataflow\_candidates.shape[1];\\

% \Comment*[r]{\textbf{\textcolor{blue}{num\_LoopOrders = candidate\_dataflows.shape[0]}}}
% \Comment*[r]{\textbf{\textcolor{blue}{num\_ClusterSzs = candidate\_dataflows.shape[1]}}}
\For{LoopOrder\_id $=$ 1 \KwTo num\_LoopOrders}{
    % candidate\_mappings\_each\_loop\_order\_with\_tile\_size = []\\
    \For{ClusterSz\_id $=$ 1 \KwTo num\_ClusterSzs}{
        % $\gamma$ $\leftarrow$ candidate\_dataflows[\textit{LoopOrder\_id}][\textit{ClusterSz\_id}][1]; DirectiveOrder $\leftarrow$ candidate\_dataflows[\textit{LoopOrder\_id}][\textit{ClusterSz\_id}][2];\\
        % outer\_LoopOrder $\leftarrow$ candidate\_dataflows[\textit{LoopOrder\_id}][\textit{ClusterSz\_id}][3]; inner\_LoopOrder $\leftarrow$ candidate\_dataflows[\textit{LoopOrder\_id}][\textit{ClusterSz\_id}][4];\\
        
        % \Comment*[r]{\textcolor{black}{$\lambda$: size of cluster}}
        
        [$\lambda$, DirectiveOrder, outer\_LoopOrder, inner\_LoopOrder] $\leftarrow$ \textbf{get\_dataflow}(\textit{LoopOrder\_id}, \textit{ClusterSz\_id}, dataflow\_candidates);\\
        
        % [$\lambda$, DirectiveOrder, outer\_LoopOrder, inner\_LoopOrder] $\leftarrow$ dataflow\_candidates[\textit{LoopOrder\_id}][\textit{ClusterSz\_id}]\Comment*[r]{\textcolor{black}{$\lambda$: size of cluster}}
        
        % \textbf{{\small //}} all\_possible\_TileSzs $\leftarrow$ \textbf{get\_non\_pruned\_Tile\_Sizes}(\hspace{10mm}{\scriptsize$\lambda$,DirectiveOrder,outer\_LoopOrder,inner\_LoopOrder,$\alpha$,$\beta$,\textit{P},\textit{M},\textit{N},\textit{K}});\\
        
        % optimal\_outer\_TileSz $\leftarrow$ \textbf{calc\_optimal\_outer\_Tile\_Size}($\gamma$, DirectiveOrder, outer\_LoopOrder, $\beta$, \textit{P}, \textit{M}, \textit{N}, \textit{K})\\
        
        [$\textit{T}^{\textit{out}}_{\textit{M}}$, $\textit{T}^{\textit{out}}_{\textit{N}}$, $\textit{T}^{\textit{out}}_{\textit{K}}$] $\leftarrow$ \textbf{calculate\_candidate\_outer\_Tile\_Size}($\lambda$, DirectiveOrder, outer\_LoopOrder, $\beta$, \textit{P}, \textit{M}, \textit{N}, \textit{K});\Comment*[r]{\textcolor{black}{Eq. \ref{eq:optimal_tile_size_L2_solution}}}
        
        % optimal\_inner\_TileSz $\leftarrow$ \textbf{calc\_optimal\_inner\_Tile\_Size}(optimal\_outer\_TileSz, $\gamma$, DirectiveOrder, inner\_LoopOrder, $\alpha$, \textit{P}, \textit{M}, \textit{N}, \textit{K})\\
        
        [$\textit{T}^{\textit{in}}_{\textit{M}}$, $\textit{T}^{\textit{in}}_{\textit{N}}$, $\textit{T}^{\textit{in}}_{\textit{K}}$] $\leftarrow$ \textbf{calculate\_candidate\_inner\_Tile\_Size}($\textit{T}^{\textit{out}}_{\textit{M}}$, $\textit{T}^{\textit{out}}_{\textit{N}}$, $\textit{T}^{\textit{out}}_{\textit{K}}$, $\lambda$, DirectiveOrder, inner\_LoopOrder, $\alpha$, \textit{P}, \textit{M}, \textit{N}, \textit{K});\Comment*[r]{\textcolor{black}{Eq. \ref{eq:optimal_tile_size_L1_solution}}}

        % [$\textit{T}^{\textit{in}}_{\textit{M}}$, $\textit{T}^{\textit{in}}_{\textit{N}}$, $\textit{T}^{\textit{in}}_{\textit{K}}$] $\leftarrow$ \textbf{calc\_candidate\_inner\_Tile\_Size}($\textit{T}^{\textit{out}}_{\textit{M}}$,$\textit{T}^{\textit{out}}_{\textit{N}}$,$\textit{T}^{\textit{out}}_{\textit{K}}$,$\lambda$,DirectiveOrder,inner\_LoopOrder,$\alpha$,\textit{P},\textit{M},\textit{N},\textit{K});\\
        
        % \Comment*[r]{\textcolor{black}{Eq. \ref{eq:optimal_tile_size_L1_solution}}}

        % \Comment*[r]{\textbf{\textcolor{black}{optimal\_outer\_TileSz $=$ [$\textit{T}^{\textit{out}}_{\textit{M}}$, $\textit{T}^{\textit{out}}_{\textit{N}}$, $\textit{T}^{\textit{out}}_{\textit{K}}$]}}\\
        % \textbf{\textcolor{black}{optimal\_inner\_TileSz $=$ [$\textit{T}^{\textit{in}}_{\textit{M}}$, $\textit{T}^{\textit{in}}_{\textit{N}}$, $\textit{T}^{\textit{in}}_{\textit{K}}$]}}
        % }
        
        % \Comment*[r]{\textbf{\textcolor{blue}{optimal\_inner\_TileSz $\leftarrow$ [$\textit{T}^{\textit{in}}_{\textit{M}}$, $\textit{T}^{\textit{in}}_{\textit{N}}$, $\textit{T}^{\textit{in}}_{\textit{K}}$]}}}\\
        
        % pruned\_TileSzs $\leftarrow$ \textbf{get\_pruned\_Tile\_Sizes}(optimal\_outer\_TileSz, optimal\_inner\_TileSz, DirectiveOrder, inner\_LoopOrder, $\alpha$, $\beta$)\\
        
        pruned\_TileSzs $\leftarrow$ \textbf{get\_pruned\_Tile\_Sizes}($\textit{T}^{\textit{out}}_{\textit{M}}$, $\textit{T}^{\textit{out}}_{\textit{N}}$, $\textit{T}^{\textit{out}}_{\textit{K}}$, $\textit{T}^{\textit{in}}_{\textit{M}}$, $\textit{T}^{\textit{in}}_{\textit{N}}$, $\textit{T}^{\textit{in}}_{\textit{K}}$, DirectiveOrder, $\alpha$, $\beta$);\\
        
        Mapping\_Candidates $\leftarrow$ \textbf{generate\_mapping\_candidates}(pruned\_TileSzs, $\lambda$, DirectiveOrder, outer\_LoopOrder, inner\_LoopOrder, \textit{M}, \textit{N}, \textit{K});\\
        
    }
}
\caption{Mapping Candidates Generator in \flash}
\label{alg:flash_setup}
\end{algorithm*}
\colonparagraph{Mapping Candidate Generation Steps} Algorithm \ref{alg:flash_setup} shows how mapping candidates are generated in \flash.
\flash generates mapping candidates given the type of accelerators, hardware parameters, and the sizes of \textit{M}, \textit{N}, and \textit{K} as inputs. Based on the accelerator chosen by the user, \flash first determines three parameters -- dataflow directive order, all feasible loop orders, and all possible cluster sizes that can be explored in the mapping candidates according to the hardware constraints described in Table \ref{table:flash_mapping_comparison} in Section \ref{subsec:modeling_spatial_accelerators}. After choosing these three parameters, \flash computes the candidate tile sizes using problem dimensions and hardware parameters. These tile sizes are used to prune a huge number of possible mappings. Finally, \flash generates the pruned mapping candidates descriptions as inputs to \newmaestro and analyzes the results of \newmaestro to choose the best \gemm mapping based on projected runtime.

\begin{table*}[ht]
\centering
\caption{The candidate tile size constraints for five different mapping schemes for different accelerators where \textit{P}, $\alpha$, $\beta$ and $\lambda$ denote total number of PEs, size of S1 buffer, size of S2 buffer and size of cluster, respectively.}
\scalebox{0.68}{
\begin{tabular}{c||c|c|c|c|c}
\thickhline
\textbf{Tile Sizes}  & \textbf{Eyeriss-style} & \textbf{NVDLA-style} & \textbf{TPU-style} & \textbf{ShiDianNao-style} & \textbf{MAERI-style}\\ \hline\hline
% T_M_out
\textbf{$\textit{T}^{\textit{out}}_{\textit{M}}$}                                         & $\textit{T}^{\textit{out}}_{\textit{M}} = \frac{\lambda M}{\textit{P}}$                                        & $1 \leq \textit{T}^{\textit{out}}_{\textit{M}} \leq \frac{\sqrt{N^{2}(\lambda +1)^{2}+2\beta \lambda}-N(\lambda +1)}{2\lambda}$                                        & $1 \leq \textit{T}^{\textit{out}}_{\textit{M}} \leq \frac{\sqrt{N^{2}(\lambda +1)^{2}+2\beta \lambda}-N(\lambda +1)}{2\lambda}$                                        & $\textit{T}^{out}_{\textit{M}} = \frac{\lambda M}{\textit{P}}$ & $1 \leq \textit{T}^{\textit{out}}_{\textit{M}} \leq \sqrt{\frac{\beta}{2}+N^{2}}-N$                                                                                                                                                       \\ \hline
% T_N_out
\textbf{$\textit{T}^{\textit{out}}_{\textit{N}}$}                                         & $1 \leq \textit{T}^{\textit{out}}_{\textit{N}} \leq \frac{\sqrt{M^{2}(\lambda +1)^{2}+2\beta \lambda}-M(\lambda +1)}{2\lambda}$ & $\textit{T}^{\textit{out}}_{\textit{N}} = \frac{\lambda N}{\textit{P}}$                                        & $\textit{T}^{\textit{out}}_{\textit{N}} = \frac{\lambda N}{\textit{P}}$                                        & $1 \leq \textit{T}^{\textit{out}}_{\textit{N}} \leq \frac{\sqrt{M^{2}(\lambda +1)^{2}+2\beta \lambda}-M(\lambda +1)}{2\lambda}$ & $\textit{T}^{\textit{out}}_{\textit{N}} = \frac{\textit{N}\textit{T}^{\textit{out}}_{\textit{K}}}{\textit{P}}$                                                                                                                      \\ \hline
% T_K_out
\textbf{$\textit{T}^{\textit{out}}_{\textit{K}}$}                                         & $1 \leq \textit{T}^{\textit{out}}_{\textit{K}} \leq \frac{\sqrt{M^{2}(\lambda +1)^{2}+2\beta \lambda}-M(\lambda +1)}{2\lambda}$                                                                               & $1 \leq \textit{T}^{\textit{out}}_{\textit{K}} \leq \frac{\sqrt{N^{2}(\lambda +1)^{2}+2\beta \lambda}-N(\lambda +1)}{2\lambda}$                                                                               & $1 \leq \textit{T}^{\textit{out}}_{\textit{K}} \leq \frac{\sqrt{N^{2}(\lambda +1)^{2}+2\beta \lambda}-N(\lambda +1)}{2\lambda}$                                        & $1 \leq \textit{T}^{\textit{out}}_{\textit{K}} \leq \frac{\sqrt{M^{2}(\lambda +1)^{2}+2\beta \lambda}-M(\lambda +1)}{2\lambda}$ & $1 \leq \textit{T}^{\textit{out}}_{\textit{K}} \leq \sqrt{\frac{\beta}{2}+N^{2}}-N$                                                                              \\ \hline \hline
% Innermost cluster
% T_M_in
\textbf{$\textit{T}^{\textit{in}}_{\textit{M}}$}                                         & $1 \leq \textit{T}^{\textit{in}}_{\textit{M}} \leq \sqrt{\frac{\alpha}{2}+(T^{out}_{K})^{2}}-T^{out}_{K}$                                                                               & $1 \leq \textit{T}^{\textit{in}}_{\textit{M}} \leq \sqrt{\frac{\alpha}{2}+(T^{out}_{K})^{2}}-T^{out}_{K}$                                                                               & $1 \leq \textit{T}^{\textit{in}}_{\textit{M}} \leq \sqrt{\frac{\alpha}{2}+(T^{out}_{K})^{2}}-T^{out}_{K}$                                        & $1 \leq \textit{T}^{\textit{in}}_{\textit{M}} \leq \sqrt{\frac{\alpha}{2}+(T^{out}_{N})^{2}}-T^{out}_{N}$      & $1 \leq \textit{T}^{\textit{in}}_{\textit{M}} \leq \sqrt{\frac{\alpha+2}{2}}-1$                                                                          \\ \hline
% T_N_in
\textbf{$\textit{T}^{\textit{in}}_{\textit{N}}$}                                         & $1 \leq \textit{T}^{\textit{in}}_{\textit{N}} \leq \sqrt{\frac{\alpha}{2}+(T^{out}_{K})^{2}}-T^{out}_{K}$                                                                               & $1 \leq \textit{T}^{\textit{in}}_{\textit{N}} \leq \sqrt{\frac{\alpha}{2}+(T^{out}_{K})^{2}}-T^{out}_{K}$                                                                               & $1 \leq \textit{T}^{\textit{in}}_{\textit{N}} \leq \sqrt{\frac{\alpha}{2}+(T^{out}_{K})^{2}}-T^{out}_{K}$                                        & $\textit{T}^{\textit{in}}_{\textit{N}} = \textit{T}^{\textit{out}}_{\textit{N}}$   &  $1 \leq \textit{T}^{\textit{in}}_{\textit{N}} \leq \sqrt{\frac{\alpha+2}{2}}-1$                                                                         \\ \hline
% T_K_in
\textbf{$\textit{T}^{\textit{in}}_{\textit{K}}$}                                         & $\textit{T}^{\textit{in}}_{\textit{K}} = \textit{T}^{\textit{out}}_{\textit{K}}$                                                                               & $\textit{T}^{\textit{in}}_{\textit{K}} = \textit{T}^{\textit{out}}_{\textit{K}}$                                                                               & $\textit{T}^{\textit{in}}_{\textit{K}} = \textit{T}^{\textit{out}}_{\textit{K}}$                                        & $1 \leq \textit{T}^{\textit{in}}_{\textit{K}} \leq \sqrt{\frac{\alpha}{2}+(T^{out}_{N})^{2}}-T^{out}_{N}$             & $\textit{T}^{\textit{in}}_{\textit{K}} = 1$                                                                \\ \thickhline
\end{tabular}
}
\label{table:all_mappings_tile_size}
\end{table*}

\colonparagraph{Tile Size Constraints} 
\flash is able to restrict the search space for different accelerators, based on the input dimension and hardware constraints. While one example derivation and the corresponding constraints for tile sizes are shown in Section \ref{sec:flash}, Table \ref{table:all_mappings_tile_size} shows all the tile sizes for all the mappings. Restricting the search space using these constraints helps \flash in choosing the near-optimal mapping.

\end{document}